\pdfoutput=1
\documentclass[acmsmall,screen,nonacm]{acmart}
\renewcommand\footnotetextcopyrightpermission[1]{}

\usepackage{graphicx}
\usepackage{placeins}
\usepackage{booktabs}
\usepackage{multirow}
\usepackage{xcolor}
\usepackage{amsmath}
\usepackage{tabularx,xltabular}
\usepackage{enumitem}
\usepackage{fvextra}
\usepackage{tcolorbox}
\setlist{itemsep=1pt,topsep=3pt}

\newcommand{\finding}[3]{\par\smallskip\noindent\fcolorbox{black!60}{black!3}{\parbox{\dimexpr\linewidth-2\fboxsep-2\fboxrule}{\textbf{Finding #1: }#2}}\par\smallskip}

\DeclareRobustCommand{\task}[1]{\texttt{#1}}
\AtBeginDocument{\pdfstringdefDisableCommands{\def\task#1{#1}}}

\AtBeginDocument{\pdfstringdefDisableCommands{\def\,{ }}}

\title[When Are Parallel Subagents Effective for Coding?]{When Are Parallel Subagents Effective for Coding? An Empirical Study on Repository Construction}

\author{Zhiheng Lyu}
\affiliation{\institution{University of Illinois Urbana-Champaign}\country{USA}}
\email{zl149@illinois.edu}
\author{Chengyu Yin}
\affiliation{\institution{University of Illinois Urbana-Champaign}\country{USA}}
\email{cyin12@illinois.edu}
\author{Lingming Zhang}
\affiliation{\institution{University of Illinois Urbana-Champaign}\country{USA}}
\email{lingming@illinois.edu}

\begin{document}

\begin{abstract}
LLM-based coding agents are increasingly applied to long-horizon software development tasks, such as building an entire program from a specification. Products such as Codex and Claude Code let an agent delegate work to subagents running in parallel. Delegation can multiply the cost of a run, yet it is unclear whether and when it produces a better program. To study this question, we compare a single agent with a main agent that delegates to subagents on repository-construction tasks from NL2Repo and ProgramBench, where ProgramBench also gives the agent an executable of the program to rebuild, the \emph{reference program}. We compare their scores at equal time and equal cost, run intervention experiments, and analyze their fine-grained behavior. We find that on randomly drawn tasks, parallel subagents raise the share of hidden tests passed by only about 1 to 5 percentage points, at up to 3.8 times the cost of a single agent. Their gain is larger on \emph{hard tasks}, where we find evidence of three mechanisms. \textbf{Self-checking}: with GPT-5.6 Sol, telling a single agent in one sentence to \emph{probe the reference program first} recovers most of this gain at a fraction of the cost. From GPT-5.6 Sol to GPT-6.1 Sol, the gain shrinks as single agents check their own work more. \textbf{Fresh context}: in extended runs, which continue until a large time budget runs out, we compare a single agent that resumes its conversation after each stop with one that periodically starts a new session with an empty conversation, as a subagent does. The fresh sessions make faster early progress in every comparison. \textbf{Modularity}: the multi-agent setup produces more modular code, with more files and shorter functions, which brings its code structure closer to the upstream source than the single-agent code. We also observe similar patterns in a smaller replication with Opus 5.5 on Claude Code. Overall, the gain from parallel subagents is associated with runs in which (i) a single agent stops checking too early, (ii) a long session keeps editing its code with little improvement, or (iii) modular code helps on the task. In other runs, subagents mainly add cost.
\end{abstract}



\maketitle

\section{Introduction}

Coding products now let an agent spend more compute at test time. The tasks assigned to coding agents have grown from implementing single functions to building entire programs over several hours. Codex, Claude Code and other products let a main agent delegate parts of such a task to parallel subagents and receive their results. Multi-agent systems of this kind have been applied to parallel search~\cite{moonshot2026k25,lan2026searchswarm}, research workflows~\cite{anthropic2025research}, solving math problems~\cite{deepmind2025deepthink,li2024moreagents}, and vulnerability discovery~\cite{fang2024hptsa}. These tasks share two properties: results are easy to verify, so a correct attempt can be selected from many, and the work is easy to decompose into pieces that need no integration. Coding is less clearly suited to such systems. Software engineering experience suggests that integrating work done in parallel is costly, because adding developers increases communication and coordination overhead~\cite{brooks1995mmm}. Anthropic reports that token usage alone explains 80\% of the performance variance of its multi-agent research system, and that most coding tasks contain fewer truly parallelizable parts than research~\cite{anthropic2025research}. A run with subagents can cost several times as much as a single agent, and it is unclear whether and when that cost raises the score on coding tasks. This matters in practice, because in Codex and Claude Code a user enables subagents with one sentence and pays the extra cost whether or not the program improves.

\begin{figure}[t]\centering
  \includegraphics[width=1.0\linewidth]{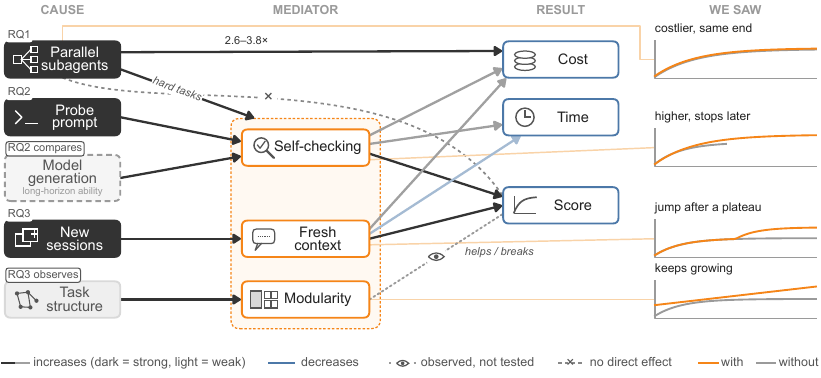}
  \caption{\label{fig:overview}\textbf{We find evidence of self-checking, fresh context and modularity behind the gain from parallel subagents. On random tasks, subagents add a small gain at several times the cost.} Cause: what we intervene on (dark) and what we compare or observe (grey). Mediator: the three factors through which parallelism could change the score. Parallel subagents raise all three on hard tasks, while modularity can also break code where modules are coupled. Result: cost, time and score. What we saw: the score-over-time curves for each cause.}\end{figure}

Repository construction, the task of building a complete program or library from a specification, is a natural setting for this question. Building software projects from scratch is now a popular use of language models, and agents are deployed to seed and grow codebases over long periods with little human oversight~\cite{yang2026programbench}. The same capability underlies reimplementing or migrating existing software, for example translating an entire repository into another language~\cite{wang2024repotransbench}. In one industrial report, 16 Claude agents working in parallel wrote a 100,000-line C compiler that builds a bootable Linux kernel~\cite{anthropic2026compiler}. Recent benchmarks ask agents to build whole repositories~\cite{zhao2025commit0,ding2025nl2repo,yang2026programbench}. These tasks require an agent to design an architecture, manage dependencies across modules and keep working for hours. Current agents rarely complete such tasks. On NL2Repo, the strongest agents pass fewer than 40\% of the tests on average~\cite{ding2025nl2repo}, and on ProgramBench no model fully resolves any task~\cite{yang2026programbench}.

In issue fixing, as in SWE-bench~\cite{jimenez2024swebench}, a fix usually changes a few lines in one place, which leaves little work to run in parallel.
Repository construction, by contrast, involves enough code to split across agents, as large software systems are split across teams. Yet it is neither easy to verify nor easy to decompose. First, the tests used for evaluation are hidden from the agent, so the agent must decide on its own when the program is complete. Second, modules share state and depend on one another, so work done in parallel must be integrated.

To understand when subagents pay off, we study the native subagents of Codex and repeat parts of the study on Claude Code (§\ref{sec:other}). We compare two setups. In \emph{Multi}, a main agent may delegate work to subagents and decides on its own whether and how to do so. In \emph{Single}, one agent works alone. We use repository-construction tasks from two benchmarks. NL2Repo~\cite{ding2025nl2repo} specifies a Python library only in natural language. ProgramBench~\cite{yang2026programbench} also provides an executable of the program to be rebuilt, the \emph{reference program}, which the agent can run on any input and compare with its own implementation. We replay every run as score-over-time and score-over-cost curves and trace the trajectory of every agent.

We ask three research questions:

\begin{itemize}
  \item \textbf{RQ1} (parallelism): Do parallel subagents change the final score, or only the time and cost?
  \item \textbf{RQ2} (self-checking): To what extent does the multi-agent gain on hard tasks come from running more checks? If a single agent is simply told to check its program more, does its score move toward the multi-agent score, and how does this change across model generations?
  \item \textbf{RQ3} (fresh context and modularity): When Single and Multi both keep working until a large budget runs out, when and why does Multi end higher at equal cost? Does restarting a single agent with an empty context move its score toward the multi-agent score? Is the remaining gap associated with how the code is divided into modules, including where that division breaks code?
\end{itemize}

\noindent Figure~\ref{fig:overview} organizes the study in four columns. The first column, \emph{Cause}, holds what we intervene on, compare or observe. In RQ1, we intervene on parallel subagents. In RQ2, we add the probe sentence to a single agent and compare three generations of the GPT Sol model family. In RQ3, we add chains of fresh sessions and observe the structure of each task. The second column, \emph{Mediator}, lists the three factors through which parallelism could change the score. \emph{Self-checking} is how much the agent checks its program before it stops, measured by how long it works and how many differential tests it runs against the reference program (§\ref{sec:rq2}). \emph{Fresh context} is whether a session starts with an empty conversation, and \emph{modularity} is whether the code is divided into modules or written as one file. The third column, \emph{Result}, holds cost, time and score. On random tasks, parallelism raises cost severalfold but raises the final score only slightly. The last column, \emph{What we saw}, sketches the score curves that we observed with and without each factor, which the three RQs examine in turn.

In \textbf{RQ1}, we vary only parallelism under ordinary use, in which each run ends when the agent declares the task complete. We draw the tasks at random from both benchmarks and run them with GPT-5.6 Sol. On ProgramBench, we repeat the comparison with GPT-6 Sol and GPT-6.1 Sol. Parallel subagents add \textbf{a small gain} in final score over a single agent, as the coordination overhead of human teams would predict. The gain is about 1 to 5 points across model generations and benchmarks, at 2.6 to 3.8 times the cost of a single agent. At equal cost, Single's score is higher than or equal to Multi's. On ProgramBench, the gain is concentrated on hard tasks. On NL2Repo, it is not, possibly because NL2Repo provides no reference program against which the extra checks of subagents can be run. RQ2 and RQ3 therefore study where the gain comes from on ProgramBench's hard tasks.

In \textbf{RQ2}, we start from how Single ends on these tasks. Single stops as soon as its own checks pass, while many hidden tests still fail. On \task{atlas}, a hard ProgramBench task that rebuilds a database schema tool, a single GPT-5.6 Sol agent stops at 47\% of the hidden tests, whereas with subagents it reaches 77\%. We therefore vary self-checking by telling a single agent, in one sentence, to probe the reference program first, and we compare three model generations to see whether the effect of probing changes with model ability. We find evidence of \textbf{self-checking}, and its effect depends on the model generation. With GPT-5.6 Sol, the probed single agent's score moves toward the multi-agent score at a fraction of the cost, and reaches 75\% on \task{atlas}. With GPT-6.1 Sol, the single agent works longer before it stops than the multi-agent runs do, possibly because newer models are better at long-horizon tasks. On \task{atlas}, it already passes 88\%, and subagents add only 2 points. As a result, on the hard tasks, the gain from subagents shrinks from GPT-5.6 Sol to GPT-6.1 Sol.

In \textbf{RQ3}, we ask what remains of the gap when neither setup may stop early. On three selected hard tasks, both setups keep working until a budget of up to 20 hours of agent time runs out. In these extended runs, Multi overtakes Single late at equal cost, unless the multi-agent code breaks. Two factors could hold Single back. The first is \emph{context rot}: as its conversation grows, Single keeps making small edits to the same code with little improvement. The second is code kept in one file. We test the first by handing Single's code to chains of fresh single-agent sessions, and examine the second by tracing how each setup divides its code into files. We find evidence of both \textbf{fresh context and modularity}. A chain of fresh sessions starts faster than a resumed session on all three tasks in both model generations, and on \task{delta} (a ProgramBench diff viewer) with GPT-6.1 Sol it also ends clearly higher. The multi-agent code is more modular, and the remaining gap varies with how Single divides its code: on \task{delta}, where Single put nearly all its code in one file, Multi ends clearly higher even at equal cost, and where Single itself wrote modules, the gap is small or reversed.

Overall, we find evidence of three mechanisms behind the gain from parallel subagents, two tested and one observed. Each is a capability that a single agent can partly acquire on its own, as newer model generations increasingly do for self-checking and modularity. We summarize our contributions as follows:
\begin{itemize}

  \item \textbf{A study of parallel subagents on randomly drawn tasks, over time and cost.} On randomly drawn tasks from NL2Repo and ProgramBench, Codex's native subagents add a small gain over a single agent at several times the cost. This holds across three generations of GPT Sol models and in a replication on Claude Code.
  \item \textbf{A mechanism analysis of the gain from subagents on hard tasks.} On hard tasks, we test self-checking and fresh context by adding each to a single agent, and we observe that the multi-agent setup writes more modular code.
  \item \textbf{Practical findings and actionable insights.} Without subagents, a single agent can recover part of the multi-agent gain in two ways. A one-sentence prompt to probe the reference program makes it check more before it stops. Restarting it in fresh sessions breaks its loop of reworking the same code, so it starts faster than a resumed session.
\end{itemize}

\section{Study Design}

\subsection{Tasks}\label{sec:tasks}
\paragraph{\textbf{Benchmark selection.}} Coding-agent benchmarks mainly fall into four categories: patching an issue in an existing repository (SWE-bench and its Verified and Pro variants~\cite{jimenez2024swebench,openai2024sweverified,deng2025swebenchpro}); terminal tasks (Terminal-Bench~\cite{merrill2026terminalbench} and Long-Horizon-Terminal-Bench~\cite{li2026lhtb}); very long tasks of mixed types (SWE-Marathon~\cite{desai2026swemarathon} and FrontierSWE~\cite{proximal2026frontierswe}); and building a whole program or library from scratch (Commit0~\cite{zhao2025commit0}, NL2Repo~\cite{ding2025nl2repo} and ProgramBench~\cite{yang2026programbench}). Our study needs three conditions: a whole repository to build, so that there is enough work to divide; hours of work; and enough hidden tests to track the score every few minutes. Only the last category meets all three. Within this category, Commit0 lets the agent run the tests it is scored on, so the agent need not judge on its own when its program is complete. We therefore select NL2Repo and ProgramBench, which hide their tests. The two differ in one respect that we expect to matter. ProgramBench provides the reference program as an executable, which the agent can run on any input to compare outputs, whereas NL2Repo gives only a written specification, as is typical of real development (Table~\ref{tab:bench}).

\begin{table}[tbp]\centering\footnotesize
\caption{\label{tab:bench}The two benchmarks. They differ mainly in that on ProgramBench the agent can run the reference program and compare its own program's output with it.}
\setlength{\tabcolsep}{5pt}
\begin{tabularx}{\linewidth}{@{}>{\raggedright\arraybackslash}p{1.75cm}>{\raggedright\arraybackslash\hsize=0.7\hsize}X>{\raggedright\arraybackslash\hsize=1.3\hsize}X@{}}
\toprule
 & \textbf{NL2Repo} & \textbf{ProgramBench} \\
\midrule
Input & A written specification of a Python library: API signatures and usage examples & The program's documentation, example inputs, and an executable of the program to be rebuilt (the \emph{reference program}) \\
\addlinespace
Output & The Python package & Source code written from scratch, plus a \texttt{compile.sh} that builds it into \texttt{./executable}, the path of the reference program \\
\addlinespace
Hidden tests & A pytest suite that imports the package and calls its functions & Each test runs the built program and compares output, exit code and written files with the reference program's \\
\addlinespace
Self-check & Only against the specification & Also by running the reference program on \emph{any input it chooses} \\
\bottomrule
\end{tabularx}
\end{table}

\paragraph{\textbf{Task selection.}} RQ1 studies ordinary use, so its tasks are drawn at random. For each benchmark, we draw 30 tasks, ten from each of the three difficulty levels. Both benchmarks rate difficulty by the size of the \emph{upstream source}, that is, the source code of the original project. ProgramBench uses its lines of code and runtime dependencies~\cite{yang2026programbench}, and NL2Repo uses its lines of code~\cite{ding2025nl2repo}. RQ2 studies the hard tasks, on which RQ1 finds the larger gain, and needs a reference program to probe. It therefore uses \emph{ProgramBench-Hard}, which consists of 16 of the 18 tasks that are rated hard in the ProgramBench release that we use. We leave out the other two tasks, \task{miller} and \task{php-src}, each with over 14,000 tests. Because we score every 5-minute snapshot rather than only the final one, we could not score \task{miller} reliably or \task{php-src} in practical time. RQ3 then lets a single agent and a multi-agent setup run past the point at which each would stop on its own. Such runs last many hours, and one such multi-agent run with GPT-5.6 Sol costs about \$500 at list price on average, compared with under \$20 for a run to the natural stop. RQ3 therefore studies three of RQ1's ProgramBench tasks in depth instead of running a larger random sample. The three tasks are \task{delta} (a diff viewer), \task{lightningcss} (a CSS minifier) and \task{solar} (a Solidity compiler). All three are hard tasks on which a single agent passed only about half of the hidden tests.

\subsection{Setups}\label{sec:setups}
\paragraph{\textbf{Runs and subagents.}} A run is one Codex CLI session in the task's container. The agent works on the files in \texttt{/workspace} with a shell and a file editor, and its first message is the benchmark's task text followed by our run prompt. In Codex, a \emph{subagent} is a separate agent that the main agent starts with a message describing its job. It runs in the same container, works on the same \texttt{/workspace}, keeps its own conversation, and reports back to the main agent when it finishes; the main agent may then send it further work. Our default model is GPT-5.6 Sol (\texttt{gpt-5.6-sol}); one setup also uses the cheaper GPT-5.6 Luna (\texttt{gpt-5.6-luna}) for its subagents. Our default benchmark is ProgramBench, since only it provides a reference program to probe. To see how the results change across model generations, we repeat RQ1's ProgramBench comparison and RQ2 with two later generations of the same model family, GPT-6 Sol (\texttt{gpt-6-sol}) and GPT-6.1 Sol (\texttt{gpt-6.1-sol}). In these runs, the same model serves as both main agent and subagent, under the same prompts and Codex version.

\begin{table}[tbp]\centering
\caption{The setups and the instructions each adds to the benchmark's task prompt.}
\label{tab:prompts}
\begin{tcolorbox}[colback=gray!4, colframe=gray!60, boxrule=0.5pt, arc=2pt, left=4pt, right=4pt, top=2pt, bottom=2pt]
\footnotesize\renewcommand{\arraystretch}{1.08}
\begin{tabularx}{\linewidth}{@{}>{\raggedright\arraybackslash}p{4.8cm} >{\raggedright\arraybackslash}p{2.0cm} >{\raggedright\arraybackslash\scriptsize}X@{}}
\toprule
\textbf{Setup} & \textbf{Instruction, placed} & \textbf{Text} \\
\midrule
\multicolumn{3}{@{}l}{\textit{Runs end at the natural stop, within a 4\,h cap; no deadline instruction}} \\
\addlinespace[1pt]
\textbf{Single} (all RQs)\newline {\scriptsize Sol works alone and starts no subagent.} & None & (the task prompt only) \\
\addlinespace[3pt]
\textbf{Multi} (RQ1, RQ2)\newline {\scriptsize Sol main agent, up to 20 subagents at once; delegation allowed, not directed. Subagents run Sol (\emph{Same Subagent}) or Luna (\emph{Weaker Subagent}); RQ1 compares both; later RQs use Same.} & \textbf{Permission sentence}, after the task prompt & Delegate independent parts of this task to Codex's native subagents when useful; you decide how many to start, how to divide the work, and how to integrate and verify their contributions. \\
\addlinespace[3pt]
\textbf{Single-Probe} (RQ2)\newline {\scriptsize Single, told to probe the reference program's interface before implementing.} & \textbf{Probe sentence}, after the task prompt & Before implementing, probe the reference executable across its whole public interface: list its commands, options and input formats, run each against the reference, and record the exact outputs and exit codes. Then implement, and check every recorded behaviour before you finish. \\
\end{tabularx}\par\nointerlineskip
\begin{tabularx}{\linewidth}{@{}>{\raggedright\arraybackslash}p{2.3cm} >{\raggedright\arraybackslash}p{1.9cm} >{\raggedright\arraybackslash\scriptsize}X@{}}
\midrule
\multicolumn{3}{@{}l}{\textit{Extended runs: continue past every stop until the time budget runs out}} \\
\addlinespace[1pt]

\textbf{Single-Extended} (RQ3)\newline {\scriptsize Single, told at every stop to continue and keep comparing against the reference program on new inputs.} & \textbf{Deadline instruction}, before the task prompt & TIME-BUDGETED EVALUATION CONTRACT (applies to the task below):\par\smallskip{}1. You have a fixed wall-clock budget of \texttt{\{hours\_remaining\}} hours starting now. When it runs out, /workspace is scored exactly as it stands at that moment, with no warning and no separate submission step. Ignore any instruction below to finish by printing COMPLETE\_\allowbreak{}TASK\_\allowbreak{}AND\_\allowbreak{}SUBMIT\_\allowbreak{}FINAL\_\allowbreak{}OUTPUT, and do not stop early. If you believe you are done, keep comparing your ./executable against the reference ./executable on new inputs, edge cases, error messages, exit codes and output formatting, and fix every difference, until the budget runs out.\par\smallskip{}2. The workspace is scored by running ./compile.sh on a fresh checkout, so a snapshot without a working ./compile.sh and ./executable scores zero no matter how much other work exists. Decide your own approach and your own order of work; whenever you do have a build, keep ./compile.sh producing a working ./executable after every change, and never leave the workspace in a state where the build is broken for long.\par\smallskip{}3. Nobody will answer questions; decide and proceed. \\
\addlinespace[3pt]
 & \textbf{Continuation text}, each time the run stops & Budget check: \texttt{\{H\}} h \texttt{\{M\}} min of the 20 h wall-clock budget remain. The workspace is NOT final. Continue: compare ./executable against the reference on new inputs, edge cases, error messages, exit codes and output formatting, and fix every difference. Do not stop. \\
\addlinespace[3pt]
\textbf{Multi-Extended} (RQ3)\newline {\scriptsize Sol main agent asked to use up to 20 Sol subagents aggressively; no permission sentence. Also gets Single-Extended's two texts.} & \textbf{Parallelism instruction}, after the deadline instruction, before the task prompt & PARALLELISM CONTRACT (speed-max):\newline Your objectives are, together: the highest possible score of /workspace when the budget ends, AND reaching that score in the least wall-clock time. Token and API cost do not matter at all; correctness and wall-clock time do. You may run up to 20 native subagents concurrently (spawn\_\allowbreak{}agent) and you are expected to use them aggressively: split the implementation into modules or feature groups that can be built and tested independently, assign each to its own subagent in parallel, and integrate their results as they return. Give every subagent a precise scope (the files it owns, the behaviours to match against the reference ./executable, and how to verify them), require it to test its own piece against the reference before returning, and never let two subagents write the same file. While subagents are working, keep integrating, building and testing yourself instead of waiting idle. \\
\bottomrule
\end{tabularx}
\par\smallskip{\scriptsize \texttt{\{hours\_remaining\}} is the run's time budget; \texttt{\{H\} h \{M\} min} is the remaining budget, and ``20 h'' is replaced by the run's budget. On Claude Code the only change is replacing ``Codex's native subagents'' with ``Claude Code's Task subagents''.}
\end{tcolorbox}
\end{table}

\paragraph{\textbf{Compared setups.}} Table~\ref{tab:prompts} defines every setup: who works and on which models, how its runs end, which RQs use it, and the exact instructions that it adds to the task prompt. Multi appends a \emph{permission sentence} to the task, which allows the main agent to delegate but does not require it. Without it, Codex forbids subagents, and the sentence is the product's documented way to enable them~\cite{openai2026codexsubagents}. Single-Probe instead appends a \emph{probe sentence}, which tells Single to probe the reference program before implementing (RQ2). Setups also differ in how a run ends. By default, a run ends at the agent's \emph{natural stop}, when it declares the task complete, within a 4-hour cap. This is how users experience the product. An \emph{extended run} instead continues past such stops, so that RQ3 can compare setups under the same budget. Each time the agent stops, it is told how much of its time budget remains and is asked to continue. When the budget runs out, the workspace is scored as it stands. Single-Extended and Multi-Extended are the extended counterparts of Single and Multi. In place of Multi's permission sentence, the main agent of Multi-Extended receives a \emph{parallelism instruction} (Table~\ref{tab:prompts}), which asks it to use up to 20 subagents aggressively, to give each subagent its own module, and to have each subagent test its module against the reference program.

\paragraph{\textbf{Other products and models.}} §\ref{sec:other} repeats parts of the study on two other products. The first is Claude Code 2.1.283 (effort high) with Opus 5.5, Grok 4.6 and MiniMax-M3. Under the same permission sentence, each of these models may spawn subagents of its own model through Claude Code's Task tool. The second is dsh, DeepSeek's open-source agent harness, with GPT-5.6 Sol. With Opus 5.5, we repeat each RQ at a smaller scale, with one run per setup and task: RQ1 on ten random tasks of each benchmark, RQ2 on \task{atlas}, and RQ3 with extended runs on \task{delta} and \task{solar}, two of the ten ProgramBench tasks.

\subsection{Runs and measures}\label{sec:measures}
\paragraph{\textbf{Codex version and reasoning level.}} RQ1 and RQ2 use Codex 0.147, with the main agent and Sol subagents at reasoning level xhigh and Luna subagents at the default level. RQ3 ran weeks later and uses Codex 0.154 at level high. Both were the product's defaults when the runs began, and we compare setups only within the same version and reasoning levels. Each \emph{cell}, that is, one setup with one model on one task, has one to three runs, and tables average a cell's runs at the same reasoning level. On ProgramBench-Hard, GPT-5.6 Sol's Single score averages three single-agent runs. Two of them had the subagent tool available, but without the permission sentence Codex forbids its use, and no subagent was started.

\paragraph{\textbf{Score, time and cost.}} The workspace is saved every 5 minutes in the runs of RQ1 and RQ2 and every 10 minutes in extended runs, and every saved snapshot is scored. The score is the share of \emph{active tests} passed. On ProgramBench, the active tests are the task's active test shards minus its official ignore list; on NL2Repo, they are the whole hidden suite. Following prior work on ProgramBench~\cite{zhan2026agensh}, we use the share of tests passed instead of the benchmark's official resolved and almost-resolved rates, because those rates are zero for nearly every run on hard tasks. Before any model run, we checked that our scorer agrees with the official ProgramBench evaluator on each task's reference program. A run's score at a given time is that of the last snapshot saved by then, and its final score is that of its final snapshot. We say that one setup \emph{ends higher} than another when its final score is higher. A snapshot whose build fails scores 0, so a drop in a run's score stays visible. After a natural-stop run has stopped, its curve stays at the final score. Time is the agent's working time since the run started. Cost is \emph{list-price cost}. Every request by the main agent and its subagents is priced at its model's list prices for input, cached input and output tokens. A cost multiple is the ratio of two setups' mean costs over tasks. GPT-6 Sol and GPT-6.1 Sol runs are priced at GPT-5.6 Sol's prices, so their cost figures compare token use rather than price. List prices per million tokens are \$4 for input, \$0.4 for cached input and \$20 for output with Sol, and \$0.2, \$0.02 and \$1.2 with Luna. %

\section{RQ1 (Parallelism): On Randomly Drawn Tasks, Subagents Add a Small Gain at Several Times the Cost}
\label{sec:rq1}
\begin{table}[tbp]\centering\footnotesize
\caption{\label{tab:rq1}\textbf{Multi (Same Subagent) costs more than Single and ends at a similar score.} Natural-stop runs of GPT-5.6 Sol on the random tasks of both benchmarks, with Sol subagents at xhigh and Luna subagents at the default level; scores in \%, mean over tasks.}
{\begin{tabular}{@{}lrrrrrr@{}}
\toprule
& \multicolumn{3}{c}{NL2Repo 30} & \multicolumn{3}{c}{ProgramBench 30} \\
\cmidrule(lr){2-4}\cmidrule(l){5-7}
& Single & \shortstack{Multi\\(Weaker)} & \shortstack{Multi\\(Same)} & Single & \shortstack{Multi\\(Weaker)} & \shortstack{Multi\\(Same)} \\
\midrule
\multicolumn{7}{@{}l}{\emph{Score at a given time}} \\
\quad 5 minutes & 16.9 & 30.3 & 47.2 & 0.0 & 0.0 & 0.0 \\
\quad 20 minutes & 64.6 & 59.6 & 70.1 & 69.4 & 69.7 & 68.9 \\
\quad Final & 67.7 & 65.3 & 70.6 & 76.2 & 74.0 & 77.4 \\
\midrule
\multicolumn{7}{@{}l}{\emph{What a run takes}} \\
\quad Mean duration (minutes) & 26.6 & 33.7 & 19.3 & 39.3 & 28.3 & 40.2 \\
\quad Mean list-price cost (USD) & 3.02 & 5.12 & 11.56 & 5.69 & 6.08 & 18.32 \\
\midrule
\multicolumn{7}{@{}l}{\emph{Score at a given cost}} \\
\quad Score at \$2 & 61.7 & 53.0 & 17.8 & 62.6 & 36.2 & 0.0 \\
\quad Score at \$3 & 67.1 & 60.1 & 31.5 & 67.6 & 50.3 & 0.0 \\
\bottomrule
\end{tabular}}
\end{table}

\begin{table}[tbp]\centering\footnotesize
\caption{\label{tab:rq1gen}\textbf{In every model generation, Multi costs several times as much as Single for a small gain in final score.} Single versus Multi (Same Subagent) on the random ProgramBench tasks in three model generations, with Sol subagents at xhigh; scores in \%, mean over tasks. The GPT-5.6 Sol columns repeat those of Table~\ref{tab:rq1}. Rows as in Table~\ref{tab:rq1}. With GPT-6.1 Sol, Multi's higher mean owes 2.5 points to \task{hush}, on which Single ends at 13.0 and Multi ends at 87.7. The median per-task difference is +1.5 points.}
{\setlength{\tabcolsep}{4pt}\begin{tabular}{@{}l rr rr rr@{}}
\toprule
& \multicolumn{2}{c}{GPT-5.6 Sol} & \multicolumn{2}{c}{GPT-6 Sol} & \multicolumn{2}{c}{GPT-6.1 Sol} \\
\cmidrule(lr){2-3}\cmidrule(lr){4-5}\cmidrule(l){6-7}
& Single & Multi & Single & Multi & Single & Multi \\
\midrule
\multicolumn{7}{@{}l}{\emph{Score at a given time}} \\
\quad 20 minutes & 69.4 & 68.9 & 66.9 & 64.0 & 61.8 & 66.6 \\
\quad Final & 76.2 & 77.4 & 69.5 & 71.0 & 79.5 & 84.2 \\
\midrule
\multicolumn{7}{@{}l}{\emph{What a run takes}} \\
\quad Mean duration (minutes) & 39.3 & 40.2 & 18.0 & 24.0 & 60.8 & 52.4 \\
\quad Mean list-price cost (USD) & 5.69 & 18.32 & 1.77 & 5.52 & 5.50 & 14.30 \\
\midrule
\multicolumn{7}{@{}l}{\emph{Score at a given cost}} \\
\quad Score at \$3 & 67.6 & 0.0 & 67.9 & 39.9 & 76.0 & 1.0 \\
\bottomrule
\end{tabular}}
\end{table}

RQ1 varies only the setup. On randomly drawn tasks, the same agent either works alone or is allowed to delegate. %
With our default model, GPT-5.6 Sol, we run Single and the two variants of Multi (Same and Weaker Subagent, Table~\ref{tab:prompts}) on these tasks of both benchmarks. Each run ends at its natural stop, and Table~\ref{tab:rq1} compares the setups by their scores at fixed times and at fixed budgets. The main agent delegates in every Multi run, starting a median of 7 to 10 subagents depending on the benchmark and the variant. Because Multi (Weaker Subagent) ends below Single on both benchmarks at a higher cost, Table~\ref{tab:rq1gen} compares only Single and Multi (Same Subagent) across the three model generations, on our default benchmark, ProgramBench.

On the random tasks, Multi (Same Subagent) ends above Single in all three model generations, by 1.2 to 4.7 points (Table~\ref{tab:rq1gen}), a small but consistent gain. It costs 2.6 to 3.2 times as much as Single (Table~\ref{tab:rq1gen}; 3.8 on NL2Repo in Table~\ref{tab:rq1}) and, except with GPT-6 Sol, does not run longer. But at equal cost, Multi does not lead Single on either benchmark.

\begin{figure}[t]\centering
  \includegraphics[width=0.74\linewidth]{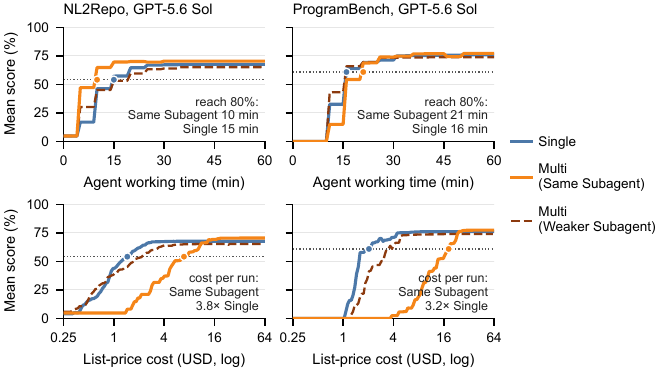}
  \caption{\label{fig:r1cost}\textbf{Over time, Multi (Same Subagent) leads early on NL2Repo but not on ProgramBench. Measured by cost, Single matches or leads both multi-agent setups.} Mean score against agent working time (top) and list-price cost (bottom), GPT-5.6 Sol on the random tasks; left: NL2Repo, right: ProgramBench. Curves come from the runs of Table~\ref{tab:rq1}. Dotted line: 80\% of Single's mean final score.}\end{figure}

\paragraph{\textbf{At equal time.}}Multi (Same Subagent) ends slightly higher after a run of similar duration, but it leads early only on NL2Repo. On NL2Repo, Multi (Same Subagent) scores higher early in the run: 47.2 against 16.9 at 5 minutes and 70.1 against 64.6 at 20 minutes (Table~\ref{tab:rq1}; Figure~\ref{fig:r1cost}, top). NL2Repo's hidden tests check each module on its own, so modules that subagents write in parallel pass tests early. On ProgramBench, by contrast, the tests run the whole program end to end. These tests can pass only once the file that connects the modules into the program exists, and Multi's main agent writes this file later than Single does (median 11.6 against 9.0 minutes). Multi therefore does not lead early on ProgramBench.

\paragraph{\textbf{At equal cost.}}When the setups are compared by cost rather than time, Multi (Same Subagent) does not lead Single until it has spent about four times Single's mean cost, and Multi (Weaker Subagent) trails Single slightly (Figure~\ref{fig:r1cost}). Single leads both multi-agent setups at almost every budget on NL2Repo, where the agent has only a written specification. On ProgramBench, Multi (Weaker Subagent), which costs about as much as Single there, trails Single at every budget, by about 2 points after \$5.

\paragraph{\textbf{Rerun variation.}}A difference between two individual runs on one task is not evidence of an effect. With GPT-5.6 Sol on the random ProgramBench tasks, two runs of the same setup on a task differ by about 7 points on average, which is more than the average gap between Single and Multi (Same Subagent). On 10 of the 30 tasks, the gain from subagents even changes sign between two repetitions of the comparison.

\begin{figure}[t]\centering
  \includegraphics[width=0.9\linewidth]{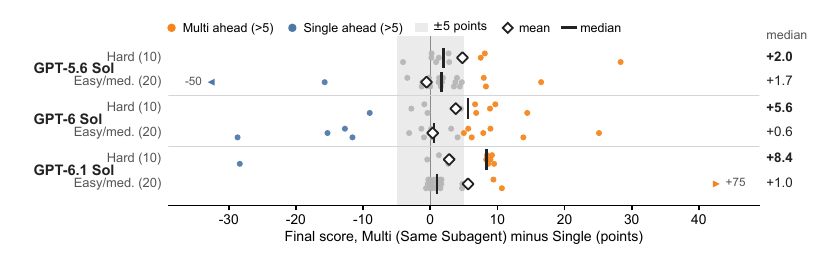}
  \caption{\label{fig:headroomgain}\textbf{Subagents gain more on hard tasks than on easy and medium ones, in all three model generations.} Per-task final score of Multi (Same Subagent) minus Single on the 30 random ProgramBench tasks, one dot per task; diamond: mean; bar: median; grey band: $\pm$5 points. Median gain on hard tasks: +2.0, +5.6 and +8.4 points with GPT-5.6, GPT-6 and GPT-6.1 Sol; on easy and medium tasks: +1.7, +0.6 and +1.0. Values beyond the axis are printed next to their dot.}\end{figure}

\paragraph{\textbf{Gain by difficulty.}}Despite this variation, a pattern appears by difficulty: subagents gain more on hard tasks. In all three model generations, the median gain on the ten hard ProgramBench tasks is larger than on the twenty easy and medium ones (Figure~\ref{fig:headroomgain}). On NL2Repo the gain is not larger on hard tasks (median +1.1 points on its hard tasks, +3.3 on the others), possibly because NL2Repo offers no reference program, and the extra checks of subagents rely on one (§\ref{sec:rq2}).

\finding{1}{On randomly drawn tasks, subagents add a small gain in final score at several times the cost of a single agent, and at equal cost a single agent matches or beats them. On ProgramBench, the gain is larger on hard tasks.}{The randomly drawn tasks of both benchmarks.}

\section{RQ2 (Self-Checking): On Hard Tasks, the Score That Subagents Add Is Associated with the Checking That Single Skips}
\label{sec:rq2}

RQ2 studies hard tasks, on which subagents gained more than on easy and medium tasks in RQ1. It focuses on self-checking, the first factor in Figure~\ref{fig:overview}. On these tasks, a setup that works longer and runs more differential tests tends to score higher, and we take these two measures to reflect self-checking. We then raise self-checking in a single agent by prompting, without subagents, and ask whether its score moves toward the multi-agent score. We use sixteen tasks of ProgramBench-Hard (§\ref{sec:tasks}), including the hard tasks of RQ1. On each, we run Single, Multi and Single-Probe to their natural stop with GPT-5.6 Sol, and repeat all three setups with GPT-6 Sol and GPT-6.1 Sol.

\begin{table}[!h]\centering\footnotesize
\caption{\label{tab:xgen}\label{tab:rq2behavior}\textbf{Multi's gain shrinks across model generations, and the probe sentence helps mainly with GPT-5.6 Sol.} RQ2 on the sixteen hard ProgramBench tasks. Score: the share of hidden tests that Single passes, and each setup's gain over Single in points; mean over tasks, median in parentheses. Run time: minutes until the natural stop. Cost: list-price cost. For run time and cost, the Single column gives Single's mean, and the Multi and Probe columns give each setup's change from Single in percent, over the tasks on which both are scored. Differential tests: commands that run the reference program and the agent's program on the same input, as the median over tasks of each task's mean over runs (§\ref{sec:rq2}). With GPT-5.6 Sol, Single-Probe is scored on 15 of the sixteen tasks; with GPT-6 Sol, Multi is scored on 15. Bold marks the values discussed in §4.}
{\setlength{\tabcolsep}{3.2pt}\begin{tabular}{@{}l rrr rrr rrr rrr@{}}
\toprule
& \multicolumn{3}{c}{Score (\%)} & \multicolumn{3}{c}{Run time} & \multicolumn{3}{c}{List-price cost} & \multicolumn{3}{c}{Differential tests} \\
\cmidrule(lr){2-4}\cmidrule(lr){5-7}\cmidrule(lr){8-10}\cmidrule(l){11-13}
& Single & Multi & Probe & Single & Multi & Probe & Single & Multi & Probe & Single & Multi & Probe \\
& & gain & gain & (min) & & & (USD) & & & & & \\
\midrule
GPT-5.6 Sol & 56.4 (54.6) & \textbf{+8.8 (+4.9)} & \textbf{+6.7 (+1.3)} & 32 & +30\% & +55\% & 5.99 & \textbf{+275\%} & \textbf{+50\%} & 15.3 & 17 & 23.5 \\
GPT-6 Sol & 55.5 (57.2) & \textbf{+6.0 (+6.7)} & +0.8 (+2.5) & 19 & +54\% & +14\% & 1.97 & +291\% & +5\% & 7 & 26 & 8 \\
GPT-6.1 Sol & 73.5 (75.2) & \textbf{+2.8 (+3.1)} & $-$1.5 ($-$0.8) & 67 & $-$11\% & $-$16\% & 6.96 & +176\% & $-$9\% & 19 & 56 & 10 \\
\bottomrule
\end{tabular}}
\end{table}

\paragraph{\textbf{Measuring self-checking.}}We define self-checking as how much an agent checks its program against the reference program before it stops, and measure it in two ways from each run's trajectory (Table~\ref{tab:xgen}). The first is the run time $T_r$, the minutes from the start of run $r$ to its natural stop. The second is the number of \emph{differential tests} $D_r$, the commands in run $r$ that run the reference program and the agent's own program on the same input. For a task $t$ with runs $R_t$, we average each measure over the runs, for example $\bar{D}_t = \frac{1}{|R_t|}\sum_{r \in R_t} D_r$, and then report the mean of $\bar{T}_t$ and the median of $\bar{D}_t$ over tasks. On these tasks, Single checks less than Multi: it runs fewer differential tests in every model generation. It also stops sooner than Multi with GPT-5.6 Sol and GPT-6 Sol; only with GPT-6.1 Sol does it run longer. Its checks are also narrow. We read by hand all 17 Single runs of GPT-5.6 Sol on seven of these tasks, and in every one of them Single stops right after a check passes. On \task{skeema}, a database tool, Single's own tests never connected to a database. On \task{gittype}, a typing-practice tool, Single compared the output of the two programs with the error stream merged into standard output. In both cases, its checks passed while many hidden tests failed.

\paragraph{\textbf{Raising self-checking in one agent.}}If the gain comes from self-checking rather than from other effects of subagents, raising self-checking in a single agent, without subagents, should move its score toward Multi's. Single-Probe tests this with one sentence that tells Single to probe the reference program before implementing. The sentence raises self-checking only with GPT-5.6 Sol, and only there does the score follow. With GPT-5.6 Sol, Single-Probe runs more differential tests than Single, stops later, and scores 6.7 points higher than Single on average, compared with 8.8 points for Multi (Table~\ref{tab:xgen}). With GPT-6 Sol, the number of differential tests barely changes, and with GPT-6.1 Sol, Single-Probe checks less than Single. With both models, Single-Probe's final score stays close to Single's (median differences +2.5 and $-$0.8 points). Either these models respond less to a one-sentence prompt, or, with GPT-6.1 Sol, Single already checks about as much as the task needs, so a reminder adds little. The second reading matches how the gain from subagents shrinks across the three model generations, from 8.8 to 6.0 to 2.8 points on average (Table~\ref{tab:xgen}).

\finding{2}{When a single agent stops while its score could still rise, the gain from subagents is associated with the checks that it skips by stopping early, and this gain can be recovered without subagents: when there is a reference program, one sentence telling GPT-5.6 Sol's single agent to probe it recovers most of the gain at a fraction of the cost. GPT-6.1 Sol's single agent works longer before it stops and leaves less room for its score to rise, so the gain from subagents shrinks.}{Sixteen ProgramBench-Hard tasks in three model generations; one to three runs per setup, final scores, and stop times from the run trajectories.}

\paragraph{\textbf{Limits of the self-checking measures.}}Run time and the number of differential tests measure how long and how often an agent checks. Neither measure shows which parts of the program the checks cover, or how the program's modules depend on each other. In extended runs, more checking can also go into the same module again and again. RQ3 shows where this leads when neither setup may stop early.

\begin{figure}[t]\centering
  \includegraphics[width=0.9\linewidth]{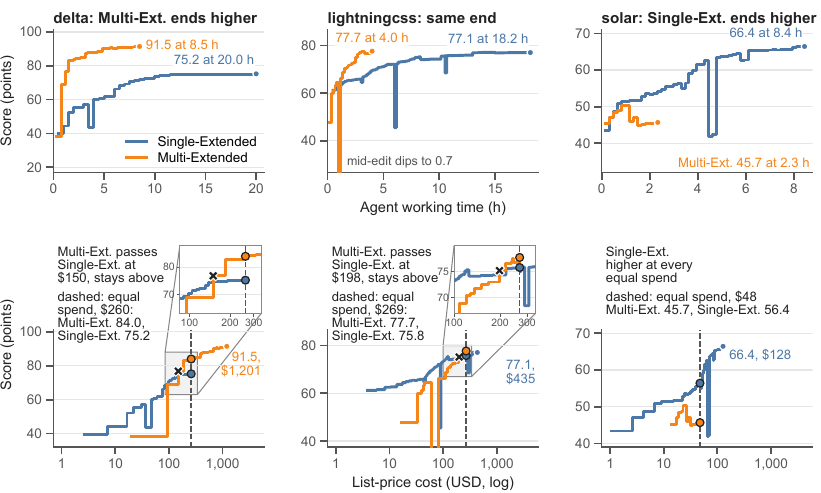}
  \caption{\label{fig:plateau}\textbf{In extended runs, Multi-Extended overtakes Single-Extended late at equal cost, except on \task{solar}, where Single-Extended ends higher.} GPT-5.6 Sol. One column per task. Top: the score of each snapshot against agent working time. Bottom: the same snapshots against list-price cost (log scale). The dashed line marks equal cost, where the cheaper run ends, and the cross marks where Multi-Extended passes Single-Extended and stays above it. On \task{delta} and \task{lightningcss} the boxed region is enlarged in the inset. Ext.\ = Extended. On \task{solar}, the main agent of the Multi-Extended run shown stopped editing at 2.3\,h. Two further runs also end below Single-Extended.}\end{figure}
\begin{figure}[t]\centering
  \includegraphics[width=0.9\linewidth]{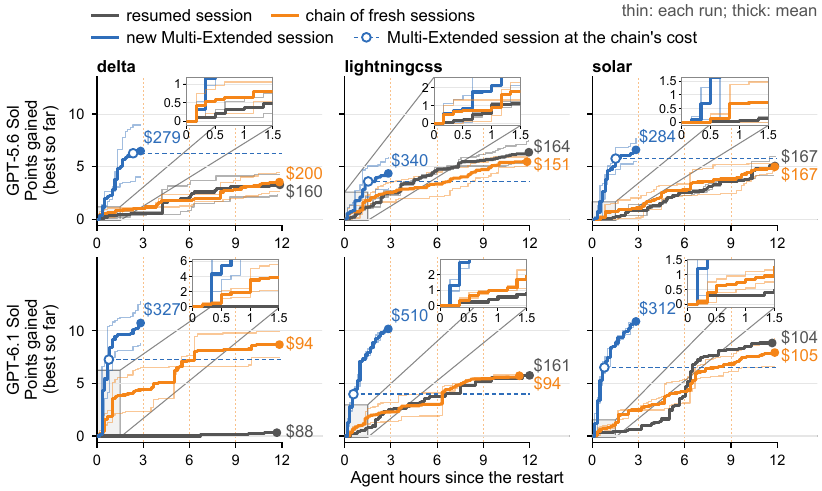}
  \caption{\label{fig:context}\textbf{A fresh session starts faster than a resumed one; later the two converge, except with GPT-6.1 Sol on \task{delta}.} Both sides start from Single-Extended's code after 8 to 12\,h. One resumes the same session, and the other is a chain of fresh sessions whose context is cleared at the dotted lines. The y-axis gives the best score so far minus the score at the start. Thin lines are single runs, thick lines are the mean, and dots mark the end of each mean line. The blue lines are a new multi-agent session (Multi-Extended) started from the same code. The open circle marks where it has spent as much as the chain, and the dashed line extends that score to 12\,h. The labels give each mean line's list-price cost since the restart. Insets magnify the shaded first 1.5\,h after the switch.}\end{figure}
\section{RQ3 (Fresh Context and Modularity): In Long Runs, the Remaining Gain Is Associated with Fresh Context and Code Structure}
\label{sec:rq3}\label{sec:rq3-stop}%

RQ3 asks whether subagents still help on very long tasks, and how. Here neither setup may stop early, so the early stop of RQ2 cannot explain a gap that remains. We test the first candidate from §1, context rot, with chains of fresh sessions (§\ref{sec:rq3-context}), and examine the second, code kept in one file, by comparing how each setup divides its code (§\ref{sec:rq3-lose}). We remove the natural stop: each time an agent stops, a user message asks it to continue, until its time budget runs out. We use three hard tasks from the random set (§2.1): \task{delta}, \task{lightningcss} and \task{solar}. Single-Extended may run up to 20\,h and Multi-Extended up to 4\,h, because Multi-Extended spends several times as much per hour, and Multi-Extended is also told to parallelize aggressively (§\ref{sec:setups}). On \task{delta}, where Multi-Extended leads clearly (below), we gave one Multi-Extended run a longer budget to see how far it would go. To save cost, a run that has passed 8\,h ends once its score has risen by less than 0.5\% of the active tests over the last 2\,h. This rule ended that \task{delta} run at 8.5\,h and Single-Extended's run on \task{solar} at 8.4\,h.

When both setups run long, Multi-Extended climbs faster once its program builds, and it is ahead at equal time. At equal cost, Single-Extended leads early. Its score then rises more slowly, and Multi-Extended overtakes it late, unless its code breaks (Figure~\ref{fig:plateau}). On \task{delta}, Multi-Extended ends clearly higher at equal cost in both model generations, and with GPT-5.6 Sol it does so in less than half of Single-Extended's time. On \task{lightningcss}, it also ends higher at equal cost in both model generations, but not significantly (§\ref{sec:rq1}). On \task{solar}, it trails Single-Extended at equal cost (§\ref{sec:rq3-lose}). This is the gap that fresh context and modularity must explain.

\subsection{Fresh Context: A New Session Starts Faster}\label{sec:rq3-context}

\paragraph{\textbf{Fresh against resumed sessions.}}A fresh session starts faster than a resumed one, and the two usually end close. \emph{Fresh context} means that an agent starts with an empty conversation on code that already exists, as every subagent does. We give it to a single agent by clearing its context every 3\,h and letting it continue from the code that it left, which we call a \emph{chain of fresh sessions}. From Single-Extended's code after 8 to 12\,h, when its score rises only slowly, we start two runs: a chain of fresh sessions, and a resumed session that relies on the product's automatic compaction. The chain is ahead over the first hour in all six panels of Figure~\ref{fig:context}. Its final score is clearly higher only with GPT-6.1 Sol on \task{delta}, where the resumed session's score barely rises. There, the resumed session spent its first hour checking only syntax highlighting, the one open item in its notes, and six hours later still passed the same tests. Fresh sessions instead read the notes and manual, ran checks over options, diffs and styles, and gained 10 to 20 tests within half an hour. In the other five panels, the two final scores are close.

\paragraph{\textbf{Restarting with subagents.}}To see whether subagents add anything beyond a fresh context, we also start a new Multi-Extended session from the same code (the blue lines in Figure~\ref{fig:context}). It \textbf{gains sooner, but not more at equal cost}. With both models, it raises the score within 3\,h, sooner than the chain, but it costs more: 1.3 to 2.2 times as much as the chain with GPT-5.6 Sol, and three to five times as much with GPT-6.1 Sol. Stopped at the chain's total cost (the open circles), it ends within rerun variation of the chain with GPT-5.6 Sol and about 1.5 points below it with GPT-6.1 Sol. From the same code, a fresh context thus gives a single agent about what subagents give at equal cost. It closes most of the gap with GPT-6.1 Sol on \task{delta}, but not with GPT-5.6 Sol, so §\ref{sec:rq3-lose} turns to how the code is divided.

\subsection{Modularity and Fragility}\label{sec:rq3-lose}

\begin{table}[tbp]\centering\footnotesize
\caption{\label{tab:structure_long}\textbf{At equal cost, Multi-Extended's code is closer to the upstream source than Single-Extended's on almost every metric, though still far from it.} Medians over the three extended-run tasks of Figure~\ref{fig:plateau} in each model generation, comparing Single-Extended's last snapshot within Multi-Extended's total cost with Multi-Extended's final snapshot. RQ3 runs only the oldest and newest of the three model generations. The Upstream column gives the upstream source. CCN is cyclomatic complexity, that is, one plus the number of branches in a function.}
\begin{tabular}{@{}lrrrrr@{}}
\toprule
 & \multicolumn{2}{c}{GPT-5.6 Sol} & \multicolumn{2}{c}{GPT-6.1 Sol} & \\
\cmidrule(lr){2-3}\cmidrule(lr){4-5}
 & Single-Ext. & Multi-Ext. & Single-Ext. & Multi-Ext. & Upstream \\
\midrule
Lines of code & 4,193 & 11,418 & 8,341 & 16,130 & 27,030 \\
Files & 1 & 20 & 21 & 26 & 99 \\
Classes & 1 & 23 & 12 & 36 & 341 \\
Functions & 113 & 443 & 331 & 616 & 1,671 \\
Lines per file & 4,193 & 1,217 & 397 & 597 & 273 \\
Mean CCN per function & 25.6 & 16.0 & 17.6 & 13.2 & 4.5 \\
Mean function length & 33 & 24 & 30 & 21 & 18 \\
Score (\%) & 75.2 & 77.7 & 92.2 & 94.0 & 100 \\
\bottomrule
\end{tabular}
\end{table}

\paragraph{\textbf{Code structure differs between setups.}}The third factor, modularity, is how finely a run divides its code into files, classes and functions. In the extended runs at equal cost (Table~\ref{tab:structure_long}), Multi-Extended's code is closer to the upstream source than Single-Extended's in both model generations on almost every metric. It has more files, classes and functions, and its functions are simpler and shorter. On \task{delta} with GPT-5.6 Sol, for example, Single-Extended wrote all 7,728 lines in one file, whereas Multi-Extended wrote 22 files, the largest of which holds 12\% of its lines. Only its files are longer with GPT-6.1 Sol (597 against 397 lines). Single agents of newer model generations also move closer to the upstream source on their own, from one file with GPT-5.6 Sol to 21 with GPT-6.1 Sol in the extended runs. Both remain far from the upstream source, which spreads its code over 99 files with a mean cyclomatic complexity of 4.5.\paragraph{\textbf{Modularity and the gain.}}Because Multi-Extended writes modules on most tasks, what varies across tasks is Single-Extended's code, and in the extended runs the remaining gap is associated with how Single-Extended divides it (Table~\ref{tab:structure_long}). On \task{delta}, Single-Extended kept nearly all its code in one file in both model generations, and it is the only task where Multi-Extended ends clearly higher. With GPT-5.6 Sol, fresh context does not explain this gap. Every single-agent path, resumed or chained, stays below Multi-Extended at equal cost by more than rerun variation. Where Single-Extended itself wrote modules (\task{lightningcss} and \task{solar} with GPT-6.1 Sol), the gap is small or negative. The exception is \task{lightningcss} with GPT-5.6 Sol, where Single-Extended also wrote one file, yet the two setups end within rerun variation.

\paragraph{\textbf{Coupling and the losses.}}We now turn to where runs lose tests that they once passed. In Figure~\ref{fig:plateau}, scores sometimes drop sharply and recover, and on \task{solar} Multi-Extended ends below Single-Extended. Tests are lost mainly where an edit breaks coupled code. In a coupling break, the agent edits code that every input passes through, usually by adding an error check, and the edited code then rejects valid inputs of other features. Among the breaks whose cause we could attribute from the diff, coupling breaks are 15 of the 22 in Multi-Extended runs and 15 of the 27 in Single-Extended runs. Multi-Extended breaks its code more often and more severely. It has more \emph{temporary breaks}, stretches of at least two snapshots in which five or more passing tests fail and most of them later pass again, and only multi-agent runs end with a program that does not build (2 of 30 extended runs, against none of 38 single-agent ones).

The agents' own checks miss these breaks. In 14 of the 15 coupling breaks in Multi-Extended runs, the agent ran its own test cases after the edit, but none of these test cases used an input that the edit now rejects. With GPT-5.6 Sol on \task{solar}, the main agent added a check that rejects a string followed by a line break, which every import line has. The 27 checks that it ran afterwards all used one-line programs (Table~\ref{tab:lost}). On this task, Multi-Extended ends below Single-Extended at equal cost. Its two reruns also end below Single-Extended, but mostly on tests that they never passed, so breaks explain only part of the gap.

\begin{table}[tbp]\centering\footnotesize
\caption{\label{tab:lost}\textbf{Multi-Extended leads on \task{delta}, where Single checks many features late or never, and trails on \task{solar}, mostly on tests that it never passed and partly on tests that its main agent broke.} Tests that pass in one setup's final snapshot and fail in the other's, GPT-5.6 Sol extended runs; the final snapshots differ in cost (Single-Extended / Multi-Extended: \task{delta} \$260 / \$1,201, \task{lightningcss} \$435 / \$269, \task{solar} \$128 / \$73); \task{solar} Multi-Extended's final snapshot at 4\,h scores 46.0. The tests that Single loses are split by when Single first compared the test's feature with the reference program. The tests that Multi-Extended loses are split by who was writing code when the test was lost, or are listed as never passed. The number of tests lost while a broad re-check reported no difference is given in parentheses. Hours: run length, Single-Extended / Multi-Extended. Net: Multi-Extended's lead in tests and points. Bold: the values behind the title.}
{\setlength{\tabcolsep}{3.5pt}
\begin{tabular}{@{}l rrrrr rrrr rr@{}}
\toprule
 & \multicolumn{5}{c}{Tests Single-Extended loses} & \multicolumn{4}{c}{Tests Multi-Extended loses} & \multicolumn{2}{c}{Net} \\
\cmidrule(lr){2-6}\cmidrule(lr){7-10}\cmidrule(l){11-12}
 & \multicolumn{4}{c}{Single-Extended first compared} & & \multicolumn{2}{c}{lost while writing} & never & & & \\
\cmidrule(lr){2-5}\cmidrule(lr){7-8}
Task (hours) & $\le$1\,h & 1--4\,h & $>$4\,h & never & All & main only & subagents too & passed & All & tests & points \\
\midrule
\task{delta} (20 / 8.5\,h) & 65 & \textbf{39} & \textbf{40} & \textbf{21} & 165 & 0 & 1 (0) & 10 & 11 & \textbf{$+$154} & $+16.2$ \\
\task{lightningcss} (18.2 / 4\,h) & 121 & 59 & 0 & 17 & 197 & 23 (0) & 22 (11) & 135 & 180 & $+17$ & $+0.6$ \\
\task{solar} (8.4 / 4\,h) & 26 & 0 & 4 & 0 & 30 & \textbf{138 (137)} & 0 & 296 & 434 & \textbf{$-$404} & $-20.4$ \\
\bottomrule
\end{tabular}}
\end{table}

On \task{lua}, an interpreter that we ran as an extra case outside our task sets (one run per setup, GPT-5.6 Sol), modules both help and hurt. Multi-Extended's main agent wrote the core and gave each of the nine standard libraries to a subagent. These libraries pass a few tests that Single-Extended's one-file code fails, but Multi-Extended ends lower overall: 69 of the 76 tests that only Single-Extended passes need the evaluator's call stack, which the debug and coroutine libraries must call back into and which no agent provided. The loss lies at an interface between modules that no agent owned.

Which factor explains the gap \textbf{depends on the task and the model generation}. With GPT-6.1 Sol on \task{delta}, a chain of fresh sessions nearly reaches Multi-Extended and a resumed session does not, so the gain is associated with fresh context. With GPT-5.6 Sol on \task{delta}, fresh context does not explain the gap, because fresh and resumed sessions end at about the same score. Only modularity is consistent with the remaining gap. On \task{lightningcss}, the two setups end within rerun variation, and on \task{solar}, Multi-Extended's code breaks.

\finding{3}{In extended runs, the multi-agent setup is ahead at equal time and overtakes a single agent late at equal cost, unless an edit breaks coupled code, and such breaks cost multi-agent runs more tests. Clearing a single agent's context gives a faster start and, on one task, closes most of the gap to the multi-agent setup. The remaining gap is associated with modularity. On \task{delta}, the one task where the multi-agent setup still ends clearly higher in both model generations, the single agent keeps its code mostly in one file.}{The three extended-run tasks with GPT-5.6 Sol and GPT-6.1 Sol (one run per setup from scratch; two Single chains and two new Multi-Extended sessions per task with GPT-5.6 Sol, two Single chains with GPT-6.1 Sol), and the 68 extended runs scored over time.}

\section{Other Products and Models}\label{sec:other}

This section asks whether the findings of RQ1--RQ3 depend on Codex and the GPT Sol models. It repeats parts of RQ1--RQ3 on Claude Code with Opus 5.5, Grok 4.6 and MiniMax-M3, and on dsh with GPT-5.6 Sol (§\ref{sec:setups}). Each cell is one run, so we read a difference as a result when it clearly exceeds rerun variation (§\ref{sec:rq1}).

\paragraph{\textbf{RQ1: the finding holds.}} With Opus 5.5 on ten random tasks of each benchmark, the gain again comes from the hard tasks, at several times the cost. On ProgramBench, Multi ends 4.7 points above Single on average (95.5 against 90.7) at 3.1 times its cost. The main agent starts 4 to 13 subagents on \task{delta}, \task{solar}, \task{jsonschema} and \task{lightningcss}. On three of these tasks, Multi ends 12.6 to 18.6 points above Single at 2.6 to 5.2 times its cost, and on \task{jsonschema} both setups score above 99. On the other six tasks, the main agent starts at most two subagents, and the setups end within 2 points. On NL2Repo, the main agent delegates in every Multi run, and Multi does not end higher. Single, Multi and Multi with Haiku~4.5 subagents end at 94.4, 90.5 and 88.0 on average, at 0.9 to 1.4 times Single's cost. The difference comes from one task, on which both multi-agent runs left out helper code that the tests import. On the other nine tasks, the three averages are within 2 points of each other (94.0, 94.5 and 93.4). Grok 4.6 and MiniMax-M3 have no natural-stop runs on random tasks.

\paragraph{\textbf{RQ2: early stopping holds; the probe sentence does not transfer to dsh or NL2Repo.}} Under dsh, GPT-5.6 Sol's Single stops as early as under Codex on four hard ProgramBench tasks (\task{atlas}, \task{broot}, \task{gittype} and \task{skeema}) and scores close to its Codex score on three of them. The probe sentence does not raise its score. On four hard NL2Repo tasks (\task{boltons}, \task{deepdiff}, \task{deslib} and \task{jinja}), GPT-5.6 Sol's Single also stops on its own, and Multi ends higher on three of the four, by 6 to 10 points. NL2Repo has no reference program to probe, so we instead tell Single to check every example in the specification. This variant ends below Single on all four tasks. Whether these models delegate at all differs. On Claude Code, under the same permission sentence, Opus 5.5 delegates on hard tasks, whereas Grok 4.6 and MiniMax-M3 never delegate on \task{delta} and \task{solar}. Like Codex's single agent, Opus 5.5's Single also stops early. On \task{atlas}, its Single stops after an hour at 81.7. When extended to 4\,h, it reaches 91.1, and when told to probe first, it reaches 91.5.

\paragraph{\textbf{RQ3: progress slows here too, the setup that ends higher differs, and fresh context is not tested.}} On Claude Code, Opus 5.5's progress also slows late in a run, as the Codex runs do. We ran it on \task{delta} and \task{solar} for 8\,h as Single and for 4\,h as Multi, continued after each stop. On \task{delta}, Single's score stays at 74.8 for almost 5\,h of its 8\,h. Unlike Codex, where Single-Extended ends higher on \task{solar}, Opus 5.5's Multi ends above its Single on both tasks, 81.3 against 78.4 on \task{delta} and 91.5 against 81.3 on \task{solar}, at about twice Single's cost (\$217 against \$94, \$167 against \$102). In six shorter pairs of runs on the same tasks, Multi is given 2\,h and Single is given 4\,h. Multi leads at 2\,h in every pair, and at its end Single is within rerun variation of Multi on average (\task{delta} 66.5 against 70.5, \task{solar} 71.4 against 69.6). We did not run chains of fresh sessions on Claude Code, so fresh context is tested on Codex alone.

\section{Discussion}\label{sec:discussion}

\paragraph{\textbf{For users.}}Each factor in Figure~\ref{fig:overview} suggests one rule. \emph{Reserve parallelism for hard tasks}, because subagents cost 2.6 to 3.8 times as much and gain only a few points (Finding 1). \emph{Get self-checking by instruction, or from a newer model}. With GPT-5.6 Sol on hard tasks, where the agent can run a reference program, one sentence that tells a single agent to probe that program moves the agent's score toward the multi-agent score, at a fraction of the multi-agent cost. With GPT-6.1 Sol, the single agent already checks its program before it stops, so neither the sentence nor subagents add much (Finding 2). \emph{Use fresh sessions when progress slows}, because a fresh session starts faster than a resumed one (Finding 3). \emph{Check how the code is divided on long tasks}. On \task{delta}, where the multi-agent setup still ended clearly higher, its code was modular, whereas the single agent put most of its code in one file (Finding 3). Last, a final score alone does not show the cost, and a gain measured on one model generation can disappear in the next (Findings 1 and 2), so setups are best compared over cost and model generation.

\paragraph{\textbf{For model and product developers.}}None of the three factors is specific to subagents, and each can be built into a single agent. Self-checking matters where results are hard to verify, the first property of §1. Single agents of newer model generations already check their own work longer, and self-checking can be trained. The target is an agent that looks for where its program is wrong rather than confirming that it runs. Fresh context needs only a product feature that clears the conversation. Modularity matters where work is hard to decompose, the second property of §1. Newer single agents also divide their code into more files, although still fewer than subagents do.  Once subagents write modules, the re-check after integration should use new inputs rather than stored ones, since a main-agent edit can break a module while the score rises (Finding 3).

\paragraph{\textbf{Open questions.}}Why a fresh context helps is still open. A session that is restarted from the same code with an empty conversation speeds up again, whereas the resumed session keeps improving slowly. This resembles a random restart that escapes a local minimum. RQ3 tests fresh context and observes modularity on three extended-run tasks. Each of the two factors gives a prediction to test on other tasks. Clearing the context when the score stops rising should help a single agent on any long task, and a single agent told to divide its code into modules should close part of the gap.

\section{Threats to Validity}\label{sec:threats}
\paragraph{\textbf{Internal validity.}}Most comparisons rest on one to three runs per setup. Multi-Extended stays clearly above Single at equal cost only on \task{delta}, with one run per setup, so we treat that comparison as a case study. Two interventions change more than the factor they target. RQ2's probe sentence also lengthens the run and raises its cost. In RQ3, Multi-Extended also receives an instruction to split the work by module and self-test each module. In a control on \task{delta}, we gave a single agent the same instruction without subagents, and the instruction accounts for at most a quarter of Multi-Extended's lead at 4\,h. The other two tasks lack this control. The edge from modularity to score is observational, and the diagram of Figure~\ref{fig:overview} organizes the mechanisms that we studied without estimating the size of any edge.

\paragraph{\textbf{Construct validity.}}Whether an agent compared a feature against the reference program is decided by fixed rules on its command text. All model generations are priced at GPT-5.6 Sol's list prices, so cost compares token use across generations.

\paragraph{\textbf{External validity.}}RQ3's tasks were chosen rather than sampled, so RQ3 identifies mechanisms and does not estimate how often they occur. RQ2's self-checking uses a reference program, which most projects lack. We study closed models, three products and only whole-program tasks, and only the compute spent within a single run. We fix reasoning effort per run type rather than sweep it, and longer reasoning is not monotonically beneficial~\cite{gema2025inverse,cuadron2025overthinking}.

\section{Related Work}\label{sec:related}

\paragraph{\textbf{LLM and coding agents.}}Coding agents are evaluated on increasingly large tasks, mostly as one agent and by final score. Agents interleave reasoning with tool calls~\cite{yao2023react,wang2024codeact}, and products embed this loop in a harness~\cite{anthropic2024effectiveagents,galster2026harness}. Benchmarks have grown from functions~\cite{chen2021codex} to repository issues~\cite{jimenez2024swebench,openai2024sweverified,deng2025swebenchpro}, machine-learning engineering~\cite{chan2025mlebench}, programs built from a specification~\cite{zhao2025commit0,ding2025nl2repo,yang2026programbench} and long tasks~\cite{kwa2025measuring,desai2026swemarathon,merrill2026terminalbench}. They are tackled by single agents~\cite{yang2024sweagent,wang2025openhands} and by a fixed pipeline at lower cost~\cite{xia2024agentless}, and trajectory studies analyze how agents work~\cite{bouzenia2025trajectories}. Codex and Claude Code now provide subagents that share the main agent's workspace~\cite{openai2026codexsubagents,anthropic2026ccsubagents}. We compare the released product with and without its subagents. With the grading tests hidden, the only feedback is the checks that the agents choose to run.

\paragraph{\textbf{Self-checking.}}LLM agents learn from feedback across attempts~\cite{shinn2023reflexion}, but without an external signal they do not reliably correct themselves~\cite{huang2024selfcorrect}. Training a model to critique flawed solutions improves its reasoning more than training it to imitate correct ones~\cite{wang2025cft}. We observe a run-time counterpart in coding agents, which confirm that their program works rather than look for its flaws. Without a provided checker, coding agents do not validate what they ship as a user would~\cite{ma2026buildingtotest}. Concurrent work finds that the final reports of agents claim review work that their transcripts do not show, even when the agents must delegate to subagents~\cite{smyth2026overclaiming}. Checking can find only the errors that the verifier can detect~\cite{liu2023evalplus,mundler2024swtbench,olausson2024selfrepair}.

\paragraph{\textbf{Test-time scaling.}}Spending more compute at inference raises scores where a verifier can select among attempts~\cite{brown2024monkeys,li2024moreagents,deepmind2025deepthink,kim2026agenticttc}, but not consistently without one~\cite{li2026generalttc}. A coding agent that builds a repository has no such verifier, so it spends extra compute within one run, for example on parallel subagents. BAGEN~\cite{lin2026bagen} shows that single agents keep consuming compute on tasks that they cannot finish. On hard repository-construction tasks, we find the opposite failure in older single agents, which use too little compute because they stop early. Concurrent work studies native subagent concurrency across products~\cite{li2026concurrency}. Samway et al.~\cite{samway2026teams} build their own team of agents and run it on all 200 ProgramBench tasks. Like us, they find that a prompt alone lets a single agent recover part of the team's gain, and that teams split the code into more files. Unlike us, they find the largest gains on easy tasks, whereas the native subagents that we study gain more on hard tasks. We follow the score over time and cost, which separates reaching a score sooner from reaching a higher one.

\paragraph{\textbf{Long sessions and fresh context.}}Concurrent work bounds the context of long coding sessions by compaction~\cite{nguyen2026cliffcompaction} or periodic forgetting~\cite{tan2026forgetting}. Agents resolve fewer tasks when they build on code written by other agents~\cite{patel2026maintainable}, and in resumed sessions they break units that they have already completed~\cite{li2026loopsbench}. In repository construction, we compare one resumed session with a chain of fresh sessions. In this setting, the structure of the inherited code, and not only the context, may limit how far a chain of fresh sessions can go.

\paragraph{\textbf{Multi-agent systems.}}Multi-agent systems based on conversation~\cite{wu2024autogen}, role-based teams~\cite{hong2024metagpt,huang2023agentcoder}, debate~\cite{du2023debate} and voting~\cite{li2024moreagents} report gains, and Anthropic's research system gains largely through higher token consumption~\cite{anthropic2025research}. In a setting close to ours, a team of parallel Claude agents built a C compiler, using GCC as the reference implementation~\cite{anthropic2026compiler}. Under matched settings the gains shrink~\cite{smit2024mad,wang2024rethinking,xu2026rethinking,fu2026more}: a single agent wins at equal thinking tokens~\cite{tran2026single}, coordination benefits diminish as the single agent strengthens~\cite{kim2026scaling}, and cost must be reported~\cite{kapoor2024agentsmatter}. Coding systems merge isolated workspaces under executable checks~\cite{geng2026caid}, compare collaboration topologies~\cite{ren2026coordination} or remove the orchestrator~\cite{zhan2026agensh}. Partitioning a repository task along its dependency graph advances the cost--score frontier over Claude Code agent teams~\cite{yang2026cocoder}, and concurrent work finds that more agents or deeper recursion do not consistently improve outcomes~\cite{yuan2026moreisless}. These studies mostly build their own systems and report final scores. We study the built-in subagents of a product, compare them with a single agent at equal time and cost over the whole run, and test two of their mechanisms by adding each to a single agent.

\paragraph{\textbf{Software engineering.}}Dividing work adds coordination and integration costs, and a reference program is a test oracle only for the inputs run on it. Adding people adds communication paths~\cite{brooks1995mmm}, and systems mirror their designers' communication structure~\cite{conway1968committees,maccormack2012mirroring}. Parallel work on one component correlates with more quality problems~\cite{perry2001parallel}, and merge conflicts~\cite{mens2002merging,ghiotto2020merge,kasi2013cassandra} also surface as build and test failures~\cite{brun2011proactive}. Another implementation can serve as a pseudo-oracle~\cite{barr2015oracle,weyuker1982nontestable}, as in back-to-back and differential testing~\cite{vouk1990backtoback,mckeeman1998differential}. These studies observe human teams and human-chosen inputs. In our setting, the main agent and its subagents choose which inputs to run on the reference program, and the score is associated more with how much a run checks against it than with the number of agents (RQ2). 

\section{Conclusion}

What does a coding agent gain from parallel subagents when it builds a whole program? On randomly drawn tasks, subagents add a small gain at several times the cost, and on hard tasks they add more. We find evidence of three mechanisms behind this gain: self-checking, fresh context and modularity. We test the first two by adding each to a single agent without subagents. With GPT-5.6 Sol on ProgramBench, one sentence that tells the single agent to probe the reference program moves its score toward the multi-agent score. A chain of fresh sessions starts faster than a resumed session once the single agent's progress has slowed. We observe the third: on \task{delta}, where subagents still end clearly higher, the single agent keeps nearly all its code in one file, while the multi-agent code is split into modules. Single agents of newer model generations check more and divide their code more on their own, so the gain from parallelism shrinks. The gain from subagents is therefore associated with runs in which (i) a single agent stops checking too early, (ii) a long session keeps reworking code that it has already written with little improvement, or (iii) modular code helps on the task. Otherwise, subagents mainly add cost.

\section*{Data Availability}
All study materials will be released in the repository \url{https://github.com/cogito233/subagent-study}: the run prompts and configurations; the replay and snapshot scorer and the scoring-correction tools; for every run used in the paper, the score of every snapshot, the pass or fail result of every hidden test in every scored snapshot, token usage per request and list-price cost; the analysis and figure scripts with the data they read; the derived trajectory data behind the analyses and the final workspaces; and the raw trajectories of these runs, with credentials, account identifiers and encrypted reasoning removed. Because the models are closed, their outputs cannot be regenerated exactly; the released data nevertheless allow every number in the paper to be recomputed without running a model. All runs used isolated containers. Coding agents assisted in writing scripts and editing text; the authors verified every number.

\bibliographystyle{ACM-Reference-Format}
\bibliography{main}

\appendix
\section{Measurement}
\label{app:measures}
This section gives the rules behind the task sets and the measures defined in §\ref{sec:measures}.

\subsection{Tasks and runs}

\paragraph{The hard tasks.} ProgramBench (commit 963063c) rates 18 tasks as hard, and we call them ProgramBench-Hard. RQ2 uses all of them except \task{miller} and \task{php-src}, which have over 14,000 tests each. \task{miller} cannot be scored on one basis across setups, because one of its test shards lost the program under test in long evaluations (§\ref{app:scoring}), and \task{php-src} cannot be evaluated in practical time. The sixteen RQ2 tasks are thus the ten hard tasks of RQ1 and \task{atlas}, \task{broot}, \task{ffmpeg}, \task{gittype}, \task{gomplate} and \task{skeema}.

\paragraph{Task sizes.} The 30 ProgramBench random tasks have 229--3,322 active tests, with a median of 819. The first 15 of them are written in Rust, except \task{doxygen}, which is written in C. Apart from \task{doxygen}, which has 481k upstream source lines, they have 0.9k--119k. On these tasks, Single scores 12--98 when we take the mean of its two runs, and it scores 5--99 in run 1, as in §\ref{sec:rq1}. Of the sixteen ProgramBench-Hard tasks of RQ2, ten are written in Rust, four in Go, and \task{doxygen} and \task{ffmpeg} in C. These tasks have 29k--615k upstream source lines, except \task{ffmpeg}, which has 2.0M, and they have 229--3,322 active tests, with a median of 1,598. GPT-5.6 Sol's Single scores 7--92 on them (Table~\ref{tab:xgen_headroom}). The 30 NL2Repo tasks have 10--1,014 hidden tests, with a median of 94.5, and Single scores 65 on average. On ProgramBench the agent's \texttt{compile.sh} builds into the path of the reference program, so the first build overwrites the reference program unless the agent has copied it first, and no test passes before a first build.

\paragraph{Runs per setup.} On the NL2Repo random tasks, Single and Multi (Weaker Subagent) run three times on each of the 30 tasks, and Multi (Same Subagent) runs once. The verifier of one drawn hard NL2Repo task, \task{pytorch-grad-cam}, did not run, so we drew the next hard task, \task{cookiecutter}. On the ProgramBench random tasks, every setup runs twice on each of the 30 tasks with GPT-5.6 Sol. The two-run comparison uses all 30 tasks. On the hard tasks with GPT-5.6 Sol, Multi and Single-Probe run twice per task, and the single-agent baseline is the mean of three runs without the permission sentence, none of which spawned a subagent (§\ref{app:rq2}). In RQ3, each setup has one or two extended runs per task, and \task{delta} has 5 Multi-Extended and 3 Single runs of 4\,h. Each takeover setup has one or two runs. NL2Repo's \task{sqlparse}, which is run with tests visible to the agent, has one Single and two Multi-Extended runs.

\paragraph{Codex versions and reasoning levels.} The two groups of runs in §\ref{sec:measures}, each of which uses one Codex version and one reasoning level, were run weeks apart. The harness for the later group (the runs whose time budget is stated in the first prompt) started with the product's defaults at the time, which were Codex 0.154 at reasoning level high, and that version and level were kept for the whole group so that its runs remain comparable. The RQ1 and RQ2 runs that were added later still used Codex 0.147 at xhigh. In RQ1 and RQ2, Sol subagents run at reasoning level xhigh, and Luna subagents run at the product's default level, medium. In the extended runs, subagents inherit the main agent's level, high. Each comparison of setups uses runs with the same levels for the main agent and for subagents. Claude Code runs use Claude Code 2.1.283 at effort high. %

\paragraph{Runtime environment.} Scores measure the product as served on the recorded dates.

\subsection{Scores, cost and delegation}

\paragraph{Scoring a snapshot.} The score definition, the snapshot reading and the rule that a snapshot whose build fails scores 0 are given in §\ref{sec:measures}. On tasks whose reference program does not itself pass every active test, scores are compared against the reference program's score. Snapshots of the NL2Repo runs and of ProgramBench run 1 were not saved during the run. Their 5-minute points are reconstructed by replaying the run's workspace edits. %

\paragraph{Snapshot score versus best-so-far.} Every run is read by the score of its own snapshots. The best-so-far score (the highest snapshot score up to a given time) is used only in the stopping rule below. Reading the best-so-far score instead changes few results. On the takeovers, the two readings differ by at most 2 points, except for one \task{delta} Multi-Extended takeover of Single's late snapshot whose final snapshot does not build and therefore scores 0, against 747 passed tests at 2.5\,h (§\ref{app:branch}). On the three extended-run tasks, the gaps at the end of the run (Multi-Extended minus Single) are $+16.2$, $+0.6$ and $-20.4$ points by snapshot score. By best-so-far score, they are $+16.2$, $+0.5$ and $-16.0$. Only \task{solar}'s Multi-Extended run ends clearly below its highest snapshot, at 46.0 against 50.4, and \task{lightningcss}'s Single run ends 0.1 points below its highest snapshot.

\paragraph{List-price cost.} Each request of each agent, whether the main agent or a subagent, is priced at the list prices of the model that served it. Uncached input, cached input and output tokens are priced separately, and output tokens include reasoning tokens. A run's cost is the sum over all of its requests. %
A subagent that starts from a copy of its parent's conversation is counted from its own first request, so the copied tokens are not counted twice. No request in any priced run exceeded 272K input tokens, the size above which list prices rise.

\paragraph{Subagent share of the code.} Each line of the final product code is credited to the last agent that added the same line in the same file, where lines are compared after whitespace normalization. The crediting uses every line that each agent added through a patch or a shell heredoc. Only source files count. Tests, probes, scratch files, documentation, examples, build outputs, comments and trivially short lines (such as a closing brace) are excluded. Crediting each line to its first writer instead of its last writer changes the share by at most 0.03.

\subsection{Edit--test cycles, comparisons and features}

\paragraph{Edit--test cycles and yield.} Each tool call of each agent is flagged as an \emph{edit} if it writes a product file (that is, a source file or \texttt{compile.sh}, but not a test, a probe or a file under \texttt{/tmp}) and as a \emph{test} if it runs the reference program, the agent's own program or the agent's tests. For each agent, an \emph{edit--test cycle} closes at the first test after at least one edit since that agent's previous test. Consecutive tests do not close a further cycle, and edits that are never followed by a test close no cycle. A call that both edits and runs a program counts as an edit followed by a test. A run's cycle count at time $t$ is the number of cycles closed by all its agents up to $t$. \emph{Yield} is the gain in best-so-far score per 100 cycles, and it is read on segments of consecutive snapshots. A segment closes once the best-so-far score has risen by 2 points or 60 cycles have passed since its start, and the first segment is merged with the second.

\paragraph{Stopping rule.} An extended run that has passed 8\,h ends once its best-so-far score has risen by less than 0.5\% of the active tests over the last 2\,h. This rule stopped \task{delta}'s longest Multi-Extended run at 8.5\,h, and it stopped \task{solar}'s Single run at 8.4\,h, while the score of that Single run was rising again.

\paragraph{Detecting tests and comparisons.} A \emph{comparison} (called a \emph{differential test} in §\ref{sec:rq2}) is a test that runs both the reference program and the agent's own program on the same input. A test that prints both outputs for the agent to read also counts, so a machine comparison such as \texttt{diff} is not required. The reference program is recognised under every path that the agent copies, moves or links it to. The agent's program is recognised as \texttt{./executable} after the agent's own build has replaced the reference, or as the agent's own source when it is run through an interpreter. Compiling, installing or running pytest is not a run of the program, nor is a mention of a program inside a file operation such as \texttt{cp} or \texttt{ls}. A script counts if its body runs both programs. Codex 0.154 can combine a patch and a command in one call, and such calls are split before detection. The detector was checked against hand labels. On 180 stratified \task{delta} commands, which come from the main agent and subagents of two Multi-Extended runs and one Single run, it reaches precision 1.00 and recall 0.96 for subagents and 1.00 for the main agent. Of 140 sampled detections on \task{solar} and \task{lightningcss}, which were split into tuning and held-out samples, 139 are true edit--test cycles. For the counts of differential tests in §\ref{sec:rq2} and Table~\ref{tab:xgen}, \texttt{./executable} counts as the reference until an event actually replaces it, such as running \texttt{compile.sh} in the workspace, or copying, moving, building or writing onto it. In these counts, the agent's program is also recognised in the argument lists of a script, such as \texttt{subprocess.run(['python3', 'atlas.py', \ldots])}. These counts also include three further cases: a check script, which is judged by what its body runs; \texttt{..{}/executable} when it is run from a subfolder of the workspace; and a run of the agent's program that is compared with an output file that the reference program wrote earlier, which is the way GPT-6 Sol mostly checks. A \texttt{compile.sh} that the shell reports as missing does not replace the reference. Of 30 sampled detections that these rules add, 29 are real comparisons. The detector works on command text, so the count is a lower bound.

\paragraph{Matching tests and commands to features.} For the RQ3 analyses of which features each setup compared (§\ref{app:gaps}), each task's hidden tests are grouped by feature: a command-line option or input format on \task{delta} (85 features); a CSS property, function, at-rule, selector or option on \task{lightningcss} (503); and a command-line option or Solidity construct on \task{solar} (98). The matching uses fixed rules, not a language model, in three steps.
(1) \emph{Test to feature.} For each hidden test we collect the command-line options that it passes, the string literals that its body uses as input and the resource files that those literals name, and we find the input formats or constructs that they contain by using a per-task vocabulary (for example, conflict markers on \task{delta} and Solidity keywords on \task{solar}). Each test is assigned one primary feature, namely the candidate whose name overlaps most with the test's module, function and parameter names. Ties go to the feature with fewer tests.
(2) \emph{Command text.} The text of a comparison is its command extended with the content of every file that it reads and that the trajectory wrote earlier (through heredocs, \texttt{printf} or \texttt{echo} redirections, or patches), so a test file that is written first and run later counts with its content.
(3) \emph{Comparison to feature.} Each feature has a set of regular expressions that must all match the command text: the long or short spelling of an option (for example \texttt{--navigate}), a pattern of an input format, or a Solidity keyword. A feature counts as \emph{compared} by time $t$ if a comparison by any agent before $t$ matches it. The match uses the command text, so a feature counts as compared even when the input did not reach the case that a hidden test exercises. The hidden tests' inputs are not visible to the agent.

\section{Scoring Corrections and Exclusions}
\label{app:scoring}
This section describes how the random task sets were drawn and lists every defect of the scoring procedure that we corrected (Table~\ref{tab:scoring}), and every run that is excluded, rerun or read as a lower bound (Table~\ref{tab:excl-runs}), each with the numbers that it affects. In total, 42 runs and 294 snapshots were rescored. %

\paragraph{Random tasks.} Each random set holds 30 tasks that were taken in a fixed random order, with ten easy, ten medium and ten hard tasks on each benchmark, and RQ1 uses all of them. Before any model run, our scorer and the official ProgramBench evaluator were run on each task's reference program, and they agree on all 30 tasks. Earlier in the draw order, \task{entr}, \task{broot} and \task{gdu} failed this check and were skipped.

\begin{table}[!htbp]\centering\footnotesize
\caption{Defects of the scoring procedure and their corrections.}
\label{tab:scoring}
\begin{tabularx}{\linewidth}{@{}>{\raggedright\arraybackslash}p{2.6cm}>{\raggedright\arraybackslash}X>{\raggedright\arraybackslash}X>{\raggedright\arraybackslash}X@{}}
\toprule
Defect & What happened & Correction & Affected numbers \\
\midrule
\task{jsonschema}: \texttt{pytest -x} & The task's official test script runs \texttt{pytest -x}, which stops at the first failure, so most tests are never run. & Snapshots are rescored without \texttt{-x}, and both versions are reported. The defect affects every snapshot. & Single 57.5 $\to$ 92.2. Tables~\ref{tab:rq1} and~\ref{tab:rq1gen} use the corrected score on every row. \\
\addlinespace
Hung tests, extended runs & One hanging test drops the remaining tests of its process. & Tests lost in this way are rerun individually and merged into the snapshot's score. & Up to 4.45 points on one \task{solar} Single snapshot (a mean of 1.2 over its snapshots). It affects all extended-run scores, including \task{solar} at 4\,h (Multi-Extended 46.0 against Single 61.4) and Figure~\ref{fig:plateau}, and the \task{solar} takeovers and chains (Tables~\ref{tab:takeover} and~\ref{tab:chain}, §\ref{app:branch}). \\
\addlinespace
Hung tests, natural-stop runs & The same loss affects run 1 of \task{solar} Single (990 tests passed after correction, 964 before) and of \task{stgit} Multi (Same Subagent) (955 after, 939 before), and run 2 of \task{solar} Multi (Same Subagent) (942 after, 922 before). On \task{jp2a}, 35 tests of 17 snapshots in six runs were still not run after this rerun and are run one per container (for example, in run 1 of Single, 452 tests passed after and 418 before). & The RQ1 rerun comparison (§\ref{app:rq1}) uses the corrected run-1 scores; Tables~\ref{tab:rq1}, \ref{tab:rq1gen} and~\ref{tab:xgen} use the corrected run-2 score. & Rerun variation (RQ1). On \task{jp2a}, it changes Single in Tables~\ref{tab:rq1} and~\ref{tab:rq1gen} by at most 0.2 points. \\
\addlinespace
NL2Repo: test isolation and denominators & Some tests fail only when the whole suite runs in one process, and the official test counts (denominators) of \task{justext} and \task{jsonlines} are swapped with each other. & Failing tests are rerun alone, file by file. \task{justext} and \task{jsonlines} use their own test counts. & Table~\ref{tab:rq1} (NL2Repo). With the official counts, the Final row reads 67.1, 65.3 and 69.4 instead of 67.7, 65.3 and 70.6, and the order of the setups is unchanged. \\
\addlinespace
NL2Repo: \task{boto} without pytest & The evaluation image of \task{boto} lacks pytest, so every snapshot of its 10 runs scored 0. & pytest is added to the image, with the same test commands and test count (1014); the reference program passes 1009 tests, and all 110 snapshots are rescored. & Table~\ref{tab:rq1} (NL2Repo) and Figure~\ref{fig:r1cost}. The Final row rises by 2.5, 1.8 and 2.6 points. \\
\addlinespace
Per-shard time limit & The ProgramBench evaluator applies an undocumented 1\,h limit to each test shard, which cuts off slow shards. & Slow shards are rescored under a relaxed limit of 4\,h per shard; runs that were not rescored are marked. & \task{skeema}, \task{miller} and \task{broot} on the hard tasks (Table~\ref{tab:xgen_runs}). \\
\addlinespace
\task{broot} with GPT-6.1 Sol & GPT-6.1 Sol's \task{broot} starts a terminal interface when \texttt{-h} is run without a terminal; the test that checks this hangs its test process, and the evaluator's per-test timeout cannot stop it, so every Single, Single-Probe and Multi snapshot after the first 25 minutes returns no result. & In the final snapshots, the tests that returned no result were rerun one at a time, and every one of them ran. & Table~\ref{tab:xgen}; §\ref{app:xgen}. \\
\task{delta}, GPT-6.1 Sol's new multi-agent sessions; GPT-5.6 Sol's Multi-Extended run 1 taking over Multi-Extended's late snapshot; GPT-6.1 Sol's Multi-Extended & In 19 tests the agent's program pipes its output into the pager \texttt{less}, which waits for a key; the per-test timeout cannot stop it, so 17 snapshots return no result (and 10 more snapshots do so with GPT-5.6 Sol). & A test that is still running the agent's program after 45\,s is stopped and counted as failed. Two snapshots that do not hang score the same under this rule. & Figure~\ref{fig:context} (GPT-6.1 Sol); §\ref{app:branch} (GPT-5.6 Sol); the GPT-6.1 Sol equal-cost gap on \task{delta} (§\ref{app:rq3}, 10.2 $\to$ 10.3). \\
\bottomrule
\end{tabularx}
\end{table}
\FloatBarrier

{\footnotesize\begin{xltabular}{\linewidth}{@{}>{\raggedright\arraybackslash}p{3.0cm}>{\raggedright\arraybackslash}X>{\raggedright\arraybackslash}X>{\raggedright\arraybackslash}X@{}}
\caption{Run-level exclusions, reruns and lower bounds.}\label{tab:excl-runs}\\
\toprule
Run & What happened & Treatment & Affected numbers \\
\midrule
\endfirsthead
\toprule
Run & What happened & Treatment & Affected numbers \\
\midrule
\endhead
\bottomrule
\endlastfoot
\multicolumn{4}{@{}l}{\emph{Random tasks (RQ1, RQ2)}} \\
ProgramBench: \task{delta}, Multi (Same Subagent) run 2 & The agent's program pipes its output into the pager \texttt{less}, which hangs the evaluator, so the run cannot be scored. & The run is rerun with the same configuration, and the rerun, which scores 53.1 after 93 minutes and costs \$30.85, replaces it. & Tables~\ref{tab:rq1gen} and~\ref{tab:xgen}, the RQ1 rerun paragraph (§\ref{app:rq1}). \\
\addlinespace
\multicolumn{4}{@{}l}{\emph{Hard tasks (RQ2)}} \\
\task{skeema}, Multi run 2 & One test shard of 542 tests, run serially, was cut off by the 1\,h per-shard limit in every snapshot. & The run is rescored under the relaxed per-shard limit and scores 83.4 (62.6 before). Both runs are used. & \task{skeema} gain $+17.0$ (two runs). \\
\addlinespace
\task{broot}, Multi run 1 & 296 tests returned no result. Every snapshot hit the evaluator's limit. & The 296 tests were rerun (234 $\to$ 406 of 538 passed). & \task{broot} gain $+14.8$. \\
\addlinespace
\task{broot}, single-agent baseline & In one of the three non-delegating runs, 272 tests returned no result. & The 272 tests were rerun (211 $\to$ 301 of 538 passed). & \task{broot} baseline 59.2. \\
\addlinespace
\task{miller}, all setups & In long evaluations one test shard lost \texttt{./executable}, and more than 1,500 tests flipped from pass to fail. The loss affects all setups alike. & \task{miller} is left out of every RQ2 comparison (§\ref{app:measures}). & RQ2 (Table~\ref{tab:xgen}). \\
\addlinespace
\multicolumn{4}{@{}l}{\emph{Extended runs and takeovers (RQ3)}} \\
\task{solar}, Multi-Extended 4\,h run & The main agent made no tool call after 141 minutes (2.3\,h) and received no continuation prompt; the code stopped changing at 160 minutes. & Readings after 2.3\,h are not drawn. & Figure~\ref{fig:plateau} (\task{solar}), §\ref{app:rq3}. \\
\addlinespace
\task{delta}, first Multi-Extended takeover of Single's late snapshot & Every snapshot after 1.7\,h times out in the evaluator, because a test hangs on the agent's program. & The takeover is dropped and replaced by two new Multi-Extended takeovers with the reference program restored. & Table~\ref{tab:takeover} (\task{delta}), §\ref{app:branch}. \\
\addlinespace
Takeovers of Single's late snapshots that overwrote the reference program & A takeover restores the handed-over snapshot of \texttt{/workspace}, but not the copy of the reference program that the original session had saved outside it. Unless the new session saves a copy first, its first build overwrites the only reference program. This affected 4 of 11 takeovers (\task{lightningcss} Single run 2; \task{solar} Single run 2 and both Multi-Extended runs), as the agents report in their trajectories (§\ref{app:branch}). & All four were rerun with the saved copy restored, and the copy was verified to stay intact throughout the run. & Takeover gains (RQ3, §\ref{app:branch}). Table~\ref{tab:takeover} gives the reruns. \\
\addlinespace
\task{delta}, takeover of Multi-Extended's late snapshot, Multi-Extended run 1 & Snapshots after 50 minutes hang the evaluator, and no snapshot exists after about 2.5\,h. & The run is rescored under the \task{delta} pager rule of Table~\ref{tab:scoring} and read at its last snapshot, at 2.5\,h, which passes 874 tests, the same as at 50 minutes. & §\ref{app:branch}. \\
\addlinespace
\multicolumn{4}{@{}l}{\emph{Other products}} \\
Claude Code runs (§\ref{app:rq5}) & GLM-5.3 on \task{solar} has no multi-agent snapshot between 55 minutes and the final workspace. Scoring of Grok 4.6's Single run on \task{delta} timed out on many snapshots. & GLM-5.3 on \task{solar} is read at its final workspace, which is scored. Grok 4.6 on \task{delta} is not reported. & Table~\ref{tab:rq5}. \\
\end{xltabular}}

\section{RQ1 Details}
\label{app:rq1}

This section gives the readings behind RQ1 (§\ref{sec:rq1}): the comparison at equal cost, the rerun variation, and the per-task scores of GPT-6.1 Sol. Which runs and tasks each row of Table~\ref{tab:rq1} uses is listed in §\ref{app:scoring}.

\paragraph{Equal cost and the early lead.} At equal cost, Single leads on NL2Repo, and on ProgramBench neither multi-agent setup leads it. On NL2Repo, Single leads both multi-agent setups at every budget from about \$1 to \$5. Below \$0.9, Multi (Weaker Subagent) is slightly ahead of Single. On ProgramBench, Single leads Multi (Same Subagent) at every budget up to about \$23. Multi (Weaker Subagent) trails Single at every budget. It is far behind up to about \$5 (36.2 against 62.6 at \$2), and it is behind by about 2 points after that. When the runs are paired over the NL2Repo tasks (with bootstrap 95\% intervals), Multi (Weaker Subagent) ends $-2.4$ [$-5.5$, $+0.4$] points from Single at 1.70 [1.50, 1.94] times Single's cost. Multi (Same Subagent) ends $+2.9$ [$-3.3$, $+8.4$] points from Single at 3.83 [3.26, 4.46] times Single's cost. The Sol main agent alone accounts for 1.6 times Single's cost in Multi (Weaker Subagent) and 1.3 times in Multi (Same Subagent). Subagents add 3\% of the total cost when they run on Luna and 67\% when they run on Sol. Over time, Multi (Same Subagent) leads Single throughout the run on NL2Repo, and Multi (Weaker Subagent) leads Single on NL2Repo only in the first 10 minutes. On ProgramBench, the mean score curve of Multi (Same Subagent) reaches 80\% of Single's mean final score after 21 minutes with GPT-6.1 Sol, against 26 minutes for Single; with GPT-5.6 Sol and GPT-6 Sol, Single reaches that score first, after 16 and 11 minutes, while Multi reaches it after 21 and 16 minutes. On NL2Repo, Multi reaches 80\% of Single's mean final score after 10 minutes and Single after 15 minutes.

\paragraph{Rerun variation.} Rerun variation is the absolute difference between the scores of two runs of one setup on a task, averaged over the 30 ProgramBench random tasks. For Single it is 7.1 points, and its median is 2.6 points. Both Single runs use the corrected scores for tests that were lost to hanging (§\ref{app:scoring}). For Multi (Same Subagent) it is 6.5 points. \task{delta}'s second Multi run cannot be scored (§\ref{app:scoring}), and a rerun with the same configuration replaces it. The two Multi runs differ in the reasoning level of the subagents. In the first run, the subagents ran at low on 15 of the tasks and at xhigh on the rest. In the second run, they ran at xhigh throughout. In the second run, as on the hard tasks of RQ2, the subagents only probe. All 30 runs delegate, to 331 subagents in total. Of these subagents, 7 wrote any workspace file, and 2, both on \task{jsonschema}, wrote a product file. Over the 30 tasks, the two multi-agent runs average $+2.2$ points above the two Single runs, at 2.6 times Single's cost. The difference is positive on 22 tasks and negative on 8. Per-task gains do not repeat across runs. For example, the gain on \task{quinn} is $+1.0$ in the first run and $+28.4$ in the second, and the gain on \task{ripsecrets} is $+2.6$ in the first run and $-15.7$ in the second. We therefore do not report any single-task gain as a result. In run 1 alone, Multi (Same Subagent) ends 3.1 points above Single at 2.0 times Single's cost. In run 2 alone, which Tables~\ref{tab:rq1} and~\ref{tab:rq1gen} show, it ends 1.2 points above Single at 3.2 times Single's cost.

\paragraph{GPT-6.1 Sol on the random tasks.} Single and Multi (Same Subagent) were repeated once each with GPT-6.1 Sol as the main agent and as the subagent model, with the same prompts, Codex version (0.147) and reasoning level as the second GPT-5.6 Sol runs (Table~\ref{tab:rq1_g61}). \task{jsonschema} is scored on the corrected basis. With GPT-6.1 Sol, Multi (Same Subagent) starts 2--7 subagents per task, against 4--20 with GPT-5.6 Sol. On \task{quinn}, Multi stops after 37 minutes with QUIC transfer still unimplemented. Its build works, but a client module that one subagent added makes every client call wait 30\,s; of the 91 tests that this affects, 37 pass when they are rerun individually. GPT-6.1 Sol's Single runs for 62 minutes on average, against 39 minutes for GPT-5.6 Sol's run 2. Multi (Same Subagent) ends 4.9 points above Single in the mean and 1.5 points above Single in the median. %

\begin{table}[tbp]\centering\small
\caption{\label{tab:rq1_g61}GPT-6.1 Sol on the 30 ProgramBench tasks of RQ1, one run per setup: final score (\%), Single's run length in minutes (in parentheses), and Multi (Same Subagent)'s subagents (the number that wrote product code / all subagents, for the first 15 tasks only). The GPT-5.6 Sol columns give Single's second run and Multi (Same Subagent)'s second run (for \task{delta}, its rerun), as in §\ref{app:rq1}.}
\begin{tabular}{@{}l rr r rr r@{}}
\toprule
 & \multicolumn{2}{c}{GPT-5.6 Sol} & & \multicolumn{3}{c}{GPT-6.1 Sol} \\
\cmidrule(lr){2-3}\cmidrule(l){5-7}
Task & Single & Multi & & Single (min) & Multi & Subagents \\
\midrule
\task{i3-style} & 94.8 & 94.8 & & 99.8 (61) & 99.4 & 2 / 3 \\
\task{dua-cli} & 92.1 & 90.8 & & 96.1 (118) & 96.1 & 2 / 2 \\
\task{delta} & 52.9 & 53.1 & & 72.6 (96) & 81.6 & 4 / 4 \\
\task{doxygen} & 36.2 & 37.1 & & 38.0 (87) & 37.6 & 5 / 5 \\
\task{datasurgeon} & 78.7 & 82.7 & & 85.5 (60) & 86.3 & 3 / 3 \\
\task{statix} & 80.9 & 88.8 & & 86.3 (54) & 96.9 & 2 / 2 \\
\task{solar} & 51.7 & 47.6 & & 63.5 (75) & 72.8 & 4 / 4 \\
\task{lightningcss} & 56.1 & 64.3 & & 71.7 (60) & 80.1 & 3 / 3 \\
\task{quinn} & 65.5 & 93.9 & & 96.7 (92) & 68.4 & 3 / 3 \\
\task{elfcat} & 93.8 & 93.6 & & 94.3 (27) & 99.1 & 3 / 3 \\
\task{tuc} & 95.1 & 96.8 & & 99.3 (73) & 99.8 & 3 / 3 \\
\task{ripsecrets} & 92.1 & 76.4 & & 91.5 (55) & 96.2 & 3 / 3 \\
\task{hex} & 98.4 & 100.0 & & 98.5 (41) & 100.0 & 2 / 3 \\
\task{stgit} & 61.4 & 64.2 & & 77.6 (93) & 86.1 & 4 / 4 \\
\task{jsonschema} & 92.1 & 94.8 & & 98.1 (64) & 99.4 & 4 / 4 \\
\task{cppcheck} & 19.6 & 23.1 & & 6.8 (36) & 7.0 & --- \\
\task{curlie} & 93.8 & 97.5 & & 85.6 (37) & 85.3 & --- \\
\task{dsq} & 84.5 & 92.8 & & 89.3 (36) & 91.0 & --- \\
\task{dutree} & 76.3 & 77.5 & & --- (40) & --- & --- \\
\task{errcheck} & 85.0 & 35.0 & & 75.3 (46) & 84.7 & --- \\
\task{figlet} & 70.6 & 87.2 & & 80.0 (35) & 80.0 & --- \\
\task{hush} & 91.1 & 89.9 & & 13.0 (48) & 87.7 & --- \\
\task{jp2a} & 89.2 & 91.1 & & 98.1 (79) & 99.4 & --- \\
\task{lazygit} & 64.7 & 66.0 & & 72.9 (58) & 75.6 & --- \\
\task{mdbook} & 83.7 & 80.3 & & 91.7 (63) & 93.2 & --- \\
\task{nnn} & 94.5 & 99.0 & & 99.2 (77) & 99.8 & --- \\
\task{oha} & 87.8 & 92.5 & & 92.9 (67) & 94.6 & --- \\
\task{pingu} & 97.1 & 98.4 & & 99.5 (22) & 99.5 & --- \\
\task{rumdl} & 54.7 & 62.2 & & 72.8 (73) & 81.4 & --- \\
\task{typst} & 50.3 & 50.5 & & 61.1 (47) & 70.6 & --- \\
\midrule
Mean & 76.2 & 77.4 & & 79.6 & 84.5 & \\
\bottomrule
\end{tabular}
\end{table}

\section{RQ2 Details}
\label{app:rq2}

This section gives the readings behind RQ2 (§\ref{sec:rq2}): the GPT-5.6 Sol runs on four hard tasks, every run of the three setups on these four tasks in three model generations, the per-task values of the sixteen RQ2 tasks, and who writes the code.

\subsection{GPT-5.6 Sol on the Hard Tasks}
\label{app:pbnew5}

\paragraph{The Single score on the hard tasks.} On each of the four hard tasks of Table~\ref{tab:xgen_runs} (\task{atlas}, \task{gittype}, \task{broot}, \task{skeema}), Single's score with GPT-5.6 Sol is the mean of three natural-stop runs without the permission sentence: one Single run and two runs that used a Multi variant's configuration but lacked the sentence. Without it, Codex does not let the main agent delegate, and none of the three runs started a subagent, so all three are single-agent runs.

\paragraph{Multi on the hard tasks.} Multi runs twice on each hard task and ends 15 to 30 points above Single. Its two runs differ by 3.5 points on average, which is less than the rerun variation of RQ1 (6.5 points; §\ref{app:rq1}). The prompt and the 4\,h cap are those of the second Multi run on RQ1's random tasks, and the subagents run at reasoning level xhigh. %
Every Multi run delegates to 7--20 subagents. Of the 151 subagents in the ten runs on these four tasks and \task{miller}, 4 wrote a workspace file, each of which was a probe input, and none wrote a line of the final product code. On \task{skeema}, one subagent per run wrote product code that the final code does not keep. Subagents account for 67--89\% of each run's cost.

\paragraph{Single-Probe.} Single-Probe runs twice per task to its natural stop (its scores are in Table~\ref{tab:xgen_runs}) and stops after 30, 63, 58 and 48 minutes on the four tasks. %

\paragraph{Per-run costs.} The mean list-price costs per run are as follows, in the order Single, Multi, Single-Probe: \task{atlas} \$2.2, \$31.0, \$3.8; \task{gittype} \$3.0, \$20.3, \$8.9; \task{broot} \$2.9, \$27.1, \$10.2; \task{skeema} \$2.9, \$36.8, \$8.8. A Multi run thus costs 7 to 14 times as much as Single on \task{atlas}, \task{gittype} and \task{skeema}, and 9 times as much on \task{broot}.

\subsection{Three Model Generations on the Hard Tasks}
\label{app:xgen}

The three RQ2 setups were repeated on the four hard tasks with GPT-6 Sol (one run each) and GPT-6.1 Sol (two runs each), with the same prompts, Codex 0.147, reasoning level xhigh for the main agent and the subagents, and with the subagents on the main agent's model. Table~\ref{tab:xgen_runs} gives every run. Table~\ref{tab:xgen} averages these four tasks together with the twelve further RQ2 tasks. GPT-6 Sol's Multi run on \task{broot} could not be scored. GPT-6.1 Sol's \task{broot} runs are scored after the hung tests are rerun (§\ref{app:scoring}). GPT-6 Sol and GPT-6.1 Sol are priced at GPT-5.6 Sol's list prices.

\paragraph{Run length and delegation.} On these four tasks, Single's natural stop moves from 10--27 minutes with GPT-5.6 Sol (three runs per task) to 16--34 minutes with GPT-6 Sol and 34--54 minutes with GPT-6.1 Sol, where one run stops after 111 minutes. Multi's runs last 33--57 minutes with GPT-5.6 Sol, 22--56 minutes with GPT-6 Sol and 34--82 minutes with GPT-6.1 Sol. GPT-5.6 Sol's main agent starts 7--20 subagents per Multi run, of which at most one writes product code, and that code makes up 0\% of the final lines. GPT-6 Sol's main agent starts 2--3 subagents per Multi run, of which 1--3 write code that makes up 19--88\% of the final lines. GPT-6.1 Sol's main agent starts 3--5 subagents per Multi run, of which 3--4 write code that makes up 42--72\% of the final lines.

\paragraph{Cost.} The mean list-price costs per run are given here in the order Single, Single-Probe, Multi. With GPT-6 Sol, they are \task{atlas} \$3.5, \$1.6, \$9.7; \task{gittype} \$0.8, \$1.3, \$9.8; and \task{skeema} \$2.0, \$1.3, \$6.8. With GPT-6.1 Sol, they are \task{atlas} \$4.7, \$5.0, \$11.7; \task{gittype} \$6.1, \$4.6, \$15.2; and \task{skeema} \$5.2, \$5.3, \$12.9. GPT-5.6 Sol's costs are given in §\ref{app:pbnew5}.

\begin{table}[tbp]\centering\small
\caption{\label{tab:xgen_runs}Every run of the three RQ2 setups on the four hard tasks, by model generation: final score (\%), runs separated by ``/''. For GPT-5.6 Sol, the Single row gives the three runs without the permission sentence (§\ref{app:pbnew5}). --- = not run or not scored. $^{h}$Scored only until the test harness hangs (§\ref{app:scoring}).}
\begin{tabular}{@{}l l rrrr@{}}
\toprule
Generation & Setup & \task{atlas} & \task{gittype} & \task{broot} & \task{skeema} \\
\midrule
GPT-5.6 Sol & Single & 50.4 / 45.0 / 46.7 & 51.0 / 52.2 / 51.7 & 55.9 / 57.4 / 64.3 & 65.3 / 67.7 / 73.1 \\
 & Single-Probe & 75.3 / 74.7 & 79.0 / 76.1 & 72.1 / 71.0 & 86.1 / 90.0 \\
 & Multi & 76.3 / 77.3 & 77.5 / 72.1 & 75.5 / 72.7 & 88.0 / 83.4 \\
\midrule
GPT-6 Sol & Single & 70.2 & 64.5 & 59.9 & 83.2 \\
 & Single-Probe & 72.8 & 66.9 & 62.6 & 82.7 \\
 & Multi & 79.1 & 84.3 & --- & 87.2 \\
\midrule
GPT-6.1 Sol & Single & 87.9 / 88.5 & 89.1 / 87.8 & 77.3 / 83.1 & 92.7 / 92.9 \\
 & Single-Probe & 88.5 / 88.3 & 87.3 / 79.8 & 84.8 / 77.7 & 91.6 / 89.3 \\
 & Multi & 89.5 / 91.4 & 91.7 / 91.6 & 85.3 / 83.8 & 95.1 / 95.0 \\
\bottomrule
\end{tabular}
\end{table}

\paragraph{All RQ2 tasks.} Table~\ref{tab:xgen_headroom} gives the values for each task of Table~\ref{tab:xgen}. RQ2 uses ProgramBench's eighteen hard tasks except \task{miller} and \task{php-src}, which could not be scored reliably or in practical time (§\ref{app:measures}): \task{atlas}, \task{broot}, \task{delta}, \task{doxygen}, \task{ffmpeg}, \task{gittype}, \task{gomplate}, \task{jsonschema}, \task{lazygit}, \task{lightningcss}, \task{quinn}, \task{rumdl}, \task{skeema}, \task{solar}, \task{stgit} and \task{typst}. On the twelve further tasks, each setup runs once per model generation, with three exceptions. First, GPT-5.6 Sol's Single on the ten RQ1 tasks averages RQ1's two runs. Second, GPT-5.6 Sol's Multi on these ten tasks uses only RQ1's second run, whose subagents ran at reasoning level xhigh like those of every other Multi run. On \task{delta}, whose second Multi run cannot be scored, a rerun with the same configuration replaces it. Third, on \task{gomplate}, GPT-5.6 Sol's first Single-Probe run stopped after 30 minutes at half of Single's score, and its rerun ended at about Single's score. Cells affected by the hanging test (§\ref{app:scoring}) that could not be corrected are left out. On \task{quinn}, Multi ends below Single with GPT-6 Sol ($-$9.0) and with GPT-6.1 Sol ($-$28.4). With GPT-6.1 Sol, Single reaches 97\% on \task{quinn}, while in Multi a client module that one subagent added makes every client call wait 30\,s. This task pulls GPT-6.1 Sol's mean gain down to +2.8, while the median gain is +3.1.

\begin{table}[t]
\centering\small
\caption{\label{tab:xgen_headroom}Each RQ2 task, sorted by GPT-5.6 Sol's Single score: per model generation, Single's final score (S, \%), Multi $-$ Single (M) and Single-Probe $-$ Single (P), in points; --- = not scored.}
{\setlength{\tabcolsep}{3.5pt}\begin{tabular}{@{}l rrr rrr rrr@{}}
\toprule
 & \multicolumn{3}{c}{GPT-5.6 Sol} & \multicolumn{3}{c}{GPT-6 Sol} & \multicolumn{3}{c}{GPT-6.1 Sol} \\
\cmidrule(lr){2-4}\cmidrule(lr){5-7}\cmidrule(l){8-10}
Task & S & M & P & S & M & P & S & M & P \\
\midrule
\task{ffmpeg} & 7.4 & +3.2 & +1.3 & 11.5 & +0.3 & +0.5 & 12.7 & +2.0 & +0.7 \\
\task{doxygen} & 35.4 & +1.7 & +1.3 & 37.6 & $-$0.9 & $-$1.3 & 38.0 & $-$0.4 & $-$0.9 \\
\task{delta} & 43.3 & +9.9 & --- & 36.6 & +9.7 & +5.2 & 72.6 & +9.0 & $-$0.8 \\
\task{atlas} & 47.4 & +29.5 & +27.6 & 70.2 & +9.0 & +2.7 & 88.2 & +2.3 & +0.2 \\
\task{typst} & 49.1 & +1.4 & +0.9 & 38.5 & +6.8 & +6.1 & 61.1 & +9.5 & +7.1 \\
\task{solar} & 50.9 & $-$3.2 & $-$1.8 & 44.0 & $-$0.4 & +2.4 & 63.5 & +9.2 & $-$3.9 \\
\task{rumdl} & 51.6 & +10.6 & +15.9 & 47.5 & +6.7 & $-$0.5 & 72.8 & +8.6 & $-$15.4 \\
\task{gittype} & 51.6 & +23.2 & +25.9 & 64.5 & +19.8 & +2.3 & 88.5 & +3.2 & $-$4.9 \\
\task{lightningcss} & 57.5 & +6.7 & +3.1 & 54.5 & +4.6 & +3.5 & 71.7 & +8.3 & $-$1.4 \\
\task{broot} & 59.2 & +14.8 & +12.3 & 59.9 & --- & +2.8 & 80.2 & +4.4 & +1.0 \\
\task{stgit} & 63.7 & +0.5 & +7.8 & 49.0 & +14.4 & +3.0 & 77.6 & +8.5 & $-$0.3 \\
\task{lazygit} & 64.1 & +1.9 & +0.9 & 64.3 & $-$2.8 & $-$6.9 & 72.9 & +2.7 & $-$0.6 \\
\task{skeema} & 68.7 & +17.0 & +19.3 & 83.2 & +4.0 & $-$0.5 & 92.8 & +2.3 & $-$2.3 \\
\task{quinn} & 75.5 & +18.4 & +1.1 & 82.6 & $-$9.0 & $-$17.6 & 96.7 & $-$28.4 & $-$1.3 \\
\task{gomplate} & 84.5 & +2.1 & $-$16.8 & 60.6 & +18.7 & +5.0 & 89.3 & +2.9 & $-$0.4 \\
\task{jsonschema} & 92.1 & +2.6 & +1.3 & 83.5 & +8.9 & +6.0 & 98.1 & +1.3 & $-$0.8 \\
\bottomrule
\end{tabular}}

\end{table}

\paragraph{Cells with fewer tasks of Table~\ref{tab:xgen}.} A cell of Table~\ref{tab:xgen} averages the tasks that are scored. Two score cells average fifteen tasks, because a run could not be scored: GPT-6 Sol's Multi gain (without \task{broot}) and GPT-5.6 Sol's Single-Probe gain (without \task{delta}). Three run-time changes average fewer tasks because a stop time is not on record: GPT-6.1 Sol's Multi change (15 tasks) and the Single-Probe changes of GPT-5.6 Sol and GPT-6 Sol (14 tasks each). GPT-5.6 Sol's Single-Probe gain includes \task{gomplate}, whose first Single-Probe run stopped after 30 minutes at half of Single's score, while its rerun ended at about Single's score. Without \task{gomplate}, the cell is +8.4. A change over Single is the setup's mean over tasks minus Single's mean, divided by Single's mean, computed on the tasks that both have.

\paragraph{Stop times and subagents.} Over the sixteen tasks, GPT-6.1 Sol's Single-Probe stops after a median of 50 minutes, against 65 minutes for Single. GPT-5.6 Sol's main agent starts a median of 12 subagents per Multi run, and GPT-6.1 Sol's main agent starts a median of four.

\paragraph{How Single stops.} The 17 GPT-5.6 Sol Single runs that we read by hand are the three single-agent runs each on \task{atlas}, \task{gittype} and \task{skeema} (§\ref{app:pbnew5}) and the two RQ1 Single runs each on \task{delta}, \task{jsonschema}, \task{lightningcss} and \task{quinn}. We read the last 5 to 14 tool calls of each run. Every run ends with a check of its own program that passes, then a commit or a build from a fresh checkout, and then the completion signal. No run stops after a failing check. On \task{skeema}, no Single run checks its program against a reachable database server. The probes that these runs make of \texttt{init} and \texttt{pull} target unreachable hosts and compare the connection errors. On \task{gittype}, all three Single runs compare the two programs with the output and error streams captured together. The reference program writes its error and not-yet-implemented messages to the error stream, while Single's program writes them to the output stream. Of the hidden tests that a Single run fails and both Multi runs pass, 56 to 75 fail on exactly that difference.

\subsection{Who Writes the Code}
\label{app:whowrites}
\begin{figure}[tbp]\centering
  \includegraphics[width=0.94\linewidth]{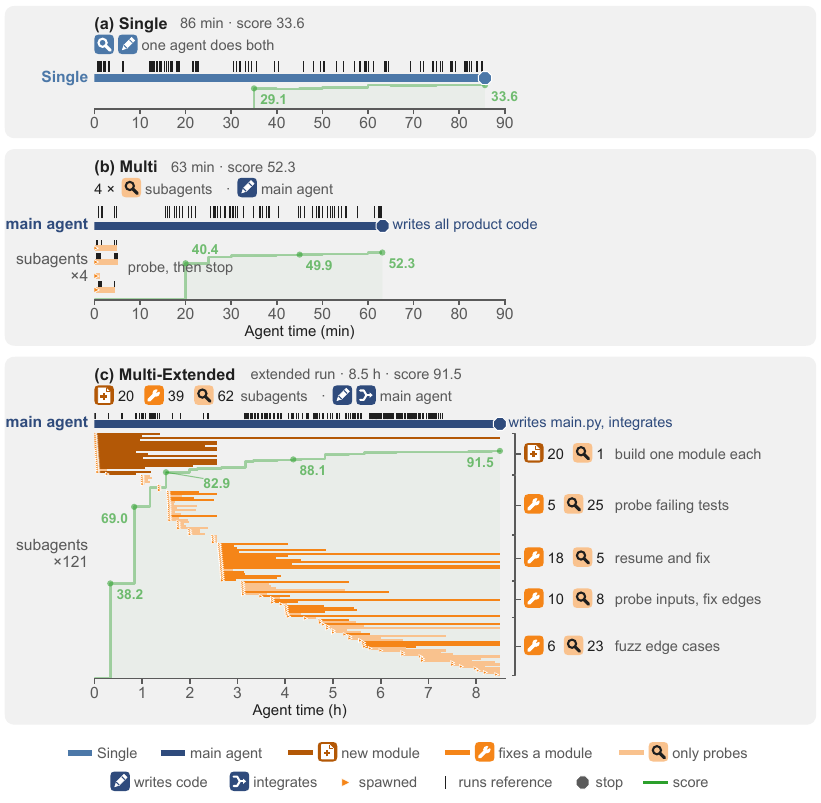}
  \caption{\label{fig:callgraph}\textbf{When it is merely permitted to delegate, the main agent sends subagents to probe the reference program and writes the code itself. When it is asked for parallelism, it has the subagents write and fix modules, and it integrates the modules.} Three GPT-5.6 Sol runs on \task{delta}: (a) Single, (b) Multi (RQ2), (c) Multi-Extended over 8.5\,h (RQ3). One row per agent. Subagent icons: new file with + = writes a new module, wrench = only fixes existing modules, magnifier = only probes; main agent: pencil = writes code, merge arrow = integrates. Green step curve: score. Ticks: runs of the reference program (in (c), only the main agent's runs are shown).}\end{figure}

Figure~\ref{fig:callgraph} shows what the subagents do in each setup on \task{delta}. %
The pattern in panel (b), in which subagents only probe, is the checking that the probe sentence gives Single at a fraction of the cost (§\ref{sec:rq2}). The pattern in panel (c) is the parallel implementation whose division into modules we examine in RQ3 (§\ref{sec:rq3-lose}).

\section{RQ3 Details}
\label{app:rq3}

This section gives the readings behind §\ref{sec:rq3}. In this section, Single-Extended is also called Single, as in Figure~\ref{fig:plateau}. %

\subsection{Outcomes of the Long Runs}
\label{app:rq3-outcomes}

\paragraph{Task choice.} \task{delta} was chosen for the extended runs because, when they began, it had the lowest Single score and the largest multi-agent gain in RQ1, so its effect size is likely larger than that of a typical task. The other two tasks were chosen by a rule that was fixed before their extended runs: the official difficulty is hard, Single's RQ1 score lies between 45 and 65, and \task{delta} is excluded. Three tasks then qualified: \task{solar} (48.7), \task{jsonschema} (57.5, the score before the scoring corrections of §\ref{app:scoring}) and \task{lightningcss} (58.9). We took \task{lightningcss}, a CSS parser and minifier, and \task{solar}, a Solidity compiler, because they are different kinds of program, and left out \task{jsonschema}, a specification-driven validator like \task{lightningcss}. The multi-agent gains of these tasks in RQ1 were small (on \task{solar} they were $-5$ and $0$ points, and on \task{lightningcss} they were $+1$ and $+2$, over the two Multi variants) and played no part in the choice.

\paragraph{A ceiling task.} We also ran one extended-run pair with GPT-5.6 Sol on \task{quinn} (a QUIC protocol library), the hard task on which Multi gained most in RQ2 (+18.4). Both setups approach the maximum. Multi-Extended reaches 96.6 at 0.8\,h and \$47. It ends at 98.9 after 8\,h and \$455. Single-Extended reaches 99.0 at 4.3\,h and \$60 and stays there. Multi-Extended is ahead at equal time, and from \$60 Single-Extended is ahead at equal cost (99.0 against about 96.7). Because both scores are already close to the maximum, the task leaves no room to see whether Multi-Extended overtakes Single late in a run, so it is not part of the RQ3 analysis.

\paragraph{Readings at equal time and cost.} The order of the setups at the end of the runs (Figure~\ref{fig:plateau}) also holds at equal time and, except at isolated snapshots, at equal cost. We compare the setups at equal time at 4\,h. On \task{delta}, the 4\,h scores are read from separate 4\,h runs that are not continued after a stop: 5 runs with the parallelism instruction and 3 runs of Single. At 4\,h, Multi-Extended leads on \task{delta} (67.9 vs.\ 54.1), trails on \task{solar} (46.0 vs.\ 61.4) and leads on \task{lightningcss} (77.7 vs.\ 69.3), where Single catches up with it only at its end (77.1 after 18.2\,h). Table~\ref{tab:rq3ends} gives the final scores and costs. At equal list-price cost, we read each setup at its last scored snapshot within each budget. On \task{delta}, Single leads up to about \$100 (70.5 vs.\ 69.0), and Multi-Extended leads from about \$150 (76.8 vs.\ 74.5, and 84.0 vs.\ 75.2 at \$260). On \task{lightningcss}, Multi-Extended leads by at most 2 points from \$200 until it ends at \$269 (77.7 vs.\ 75.8). On \task{solar}, Single leads at every budget. It leads 56.4 to 45.7 at \$48, where Multi-Extended's main agent stopped, and it stands at 66.4 from \$125 until it ends at \$128. A few snapshots taken in the middle of an edit score far below their neighbours (\task{lightningcss} Multi-Extended 0.7 at 1.0\,h; \task{lightningcss} Single 45.5 at 6.0\,h and 68.5 at 10.5\,h; \task{solar} Single 41.9 and 42.3 at 4.4--4.6\,h), and only at such snapshots does the order at equal cost flip. With GPT-6.1 Sol, Single-Extended runs 20 to 47\,h to spend as much as Multi-Extended spends in 4\,h, and Multi-Extended then ends 10.3 points higher on \task{delta} and 1.4 points higher on \task{lightningcss}. With GPT-5.6 Sol, Multi-Extended leads Single by 8.7 points at equal cost on \task{delta}, and from 12 to 20\,h Single gains half a point on \task{delta}.

\paragraph{Breaks in the extended runs.} Multi-Extended has more \emph{temporary breaks} (§\ref{sec:rq3-lose}), stretches of at least two snapshots in which five or more passing tests fail and most of them later pass again. With GPT-5.6 Sol a temporary break happens 0.32 times per hour for Multi-Extended against 0.27 times for Single-Extended, and with GPT-6.1 Sol it happens 0.33 times per hour against 0.05 times. With GPT-5.6 Sol, Multi-Extended also loses about ten times as many tests for good per hour as Single-Extended (14.4 against 1.4). Most of these breaks do not involve subagent code: only 6 of Multi-Extended's 15 coupling breaks come from editing or calling a subagent's code, and they cost 20 of the 359 tests that its runs passed and then never passed again. Single-Extended's score also drops at times (Figure~\ref{fig:plateau}), but less often. With GPT-5.6 Sol on \task{solar}, the tests that Multi-Extended's main agent broke were still failing when it stopped at 2.3\,h; two further Multi-Extended runs on \task{solar}, which lost fewer tests for good, end at 54 and 57 against Single's late score of 63 to 65. With GPT-6.1 Sol on \task{solar}, Multi-Extended is still gaining when its 4\,h budget ends, and Single-Extended needs 20\,h to spend the same \$181. At equal time (4\,h) with GPT-6.1 Sol on \task{solar}, Multi-Extended stands at 85.4 against Single-Extended's 73.0, and at equal cost Single-Extended ends 6.7 points higher.

\begin{table}[tbp]\centering\small
\caption{Final score and list-price cost of the extended runs. On \task{solar}, Multi-Extended's highest snapshot (50.4, at 1\,h) is not its final one.}\label{tab:rq3ends}
\begin{tabular}{@{}lrrrrrr@{}}
\toprule
 & \multicolumn{3}{c}{Single} & \multicolumn{3}{c}{Multi-Extended} \\
\cmidrule(lr){2-4}\cmidrule(lr){5-7}
Task & Score & Hours & Cost & Score & Hours & Cost \\
\midrule
\task{delta} & 75.2 & 20.0 & \$260 & 91.5 & 8.5 & \$1,201 \\
\task{lightningcss} & 77.1 & 18.2 & \$435 & 77.7 & 4.0 & \$269 \\
\task{solar} & 66.4 & 8.4 & \$128 & 46.0 & 4.0 & \$73 \\
\bottomrule
\end{tabular}
\end{table}

\paragraph{Visible tests.} When the agent can run part of the grading tests, Single closes the gap. On NL2Repo's \task{sqlparse}, which has no reference program, we placed 93 of the 461 hidden tests in the workspace and ran Single and Multi-Extended for 4\,h. When the runs are scored on the other 368 tests, Single reaches 86.7 and the two Multi-Extended runs reach 86.1 and 87.0. On the same 368 tests, in runs without visible tests, Single (two runs) scores 71.5 and Multi-Extended (four runs) scores 82.0. A new source of feedback removes the difference without more agents.

\paragraph{Yield per cycle.} Yield (§\ref{app:measures}) counts the points that each further 100 edit--test cycles add toward a new best score. On \task{delta}, Single earns 0--0.7 points per 100 cycles once its score is above 74, while Multi-Extended still earns about 12 points per 100 cycles until its score reaches 83. On \task{solar}, Single's late gains come from adding functions. Among the intervals between consecutive snapshots, the 13 intervals in which it added functions add a net 14.8 points, and the other 35 intervals add a net 6.4 points. Its final comparison against the reference program reports 6 mismatches, while about 600 of its 664 failing tests pass in no other run. %

\paragraph{The Multi-Extended instruction on \task{delta}.} The parallelism instruction of Multi-Extended accounts for little of its lead on \task{delta}. In the 4\,h runs that are not continued after a stop, a single agent that is given that instruction but no subagent tool averages 57.6 over 3 runs. Single averages 54.1, and the runs with subagents and the instruction average 67.9, so the instruction alone accounts for at most 3.5 of the 13.8-point lead.

\subsection{Code Structure at the Natural Stop}
\label{app:structure}

Table~\ref{tab:structure} gives the metrics of Table~\ref{tab:structure_long} for the natural-stop runs of the 20 ProgramBench tasks run in all three model generations. All metrics count product code only, and they exclude tests and vendored or generated files. Functions, classes and cyclomatic complexity are measured with lizard, and each cell is the median over tasks of one value per task. Every agent writes far denser code than the upstream sources, which spread their code over about 80 files with a mean cyclomatic complexity near 4. Across tasks, the score difference between Multi and Single correlates with their ratio of lines at 0.37 and with their ratio of files at 0.10. In Table~\ref{tab:structure_long}, Single-Extended at equal cost is read at its last snapshot whose list-price cost does not exceed Multi-Extended's total, as in Figure~\ref{fig:plateau}. With GPT-5.6 Sol on \task{delta}, Single-Extended's whole run costs less than Multi-Extended's run, so Single-Extended is read at its end, and Multi-Extended, when it is read at the same cost, is also closer to the upstream source on every metric. Multi-Extended with GPT-5.6 Sol on \task{solar} is read at the 2.3\,h cut of Figure~\ref{fig:plateau}. At equal score, Single-Extended is read at its first snapshot that reaches Multi-Extended's final score. In four of the six pairs, Single-Extended never reaches that score, and its best snapshot is used. On \task{delta}, Multi-Extended's final code has 22 files with GPT-5.6 Sol and 27 with GPT-6.1 Sol.

\begin{table}[tbp]\centering\small
\caption{\label{tab:structure}\textbf{Code written by subagents is split into more files and classes, whereas a single agent writes one large file.} Final code of the natural-stop runs on the 20 ProgramBench tasks run in all three model generations (RQ1's 15 tasks and \task{atlas}, \task{gittype}, \task{skeema}, \task{gomplate}, \task{ffmpeg}), one value per task, median over tasks; the Upstream column gives the upstream source. CCN: cyclomatic complexity, one plus the number of branches in a function.}
\begin{tabular}{@{}lrrrrrrr@{}}
\toprule
 & \multicolumn{2}{c}{GPT-5.6 Sol} & \multicolumn{2}{c}{GPT-6 Sol} & \multicolumn{2}{c}{GPT-6.1 Sol} & \\
\cmidrule(lr){2-3}\cmidrule(lr){4-5}\cmidrule(lr){6-7}
 & Single & Multi & Single & Multi & Single & Multi & Upstream \\
\midrule
Lines of code & 798 & 1,076 & 406 & 781 & 1,738 & 2,563 & 13,610 \\
Files & 1 & 1 & 1 & 4 & 3 & 7 & 78 \\
Classes & 4 & 4 & 1 & 2 & 5 & 8 & 94 \\
Functions & 39 & 40 & 17 & 36 & 64 & 112 & 668 \\
Lines per file & 714 & 936 & 406 & 305 & 410 & 463 & 196 \\
Mean CCN per function & 8.8 & 9.9 & 13.5 & 11.0 & 11.5 & 8.9 & 3.8 \\
Mean function length & 20 & 20 & 23 & 22 & 21 & 19 & 17 \\
Score (\%) & 73.7 & 85.7 & 72.2 & 79.2 & 88.9 & 91.1 & 100 \\
\bottomrule
\end{tabular}
\end{table}

\subsection{Takeovers from Late Snapshots}
\label{app:branch}
A \emph{takeover} hands the code of a late snapshot to a new session; the new Single takeovers are the first sessions of the chains of fresh sessions (Table~\ref{tab:chain}). A takeover tests whether the difference between the setups late in a run lies in the code reached so far or in how the agents check it. We hand the code of a late snapshot, first Single's and then Multi-Extended's, to new Single and new Multi-Extended sessions for 3\,h each (Table~\ref{tab:takeover}). Each session is read at its last snapshot within the 3\,h, and a snapshot whose build fails scores 0. The snapshots taken from Single come from where its progress had slowed: in the 3\,h before the takeover, the original run had gained 1.0--2.4 points, and when it was left running, it gained 0.4 points on \task{delta} and 1.5 points on \task{lightningcss} over about 8 more hours (\task{solar}'s run ended at its budget).

\paragraph{Takeovers from Single's late snapshots.} Table~\ref{tab:takeover} gives every valid takeover; reruns and replacements are listed in Table~\ref{tab:excl-runs}. %
When they took over Single's code, which was written as one file, both Multi-Extended reruns on \task{solar} once dropped at least 62 tests between snapshots and recovered them later. This is another instance of an edit to shared code that breaks what other parts of the code rely on. On \task{delta}, one Multi-Extended takeover reached 747 passing tests at 2.5\,h, then merged a subagent's commit from the subagent's own git worktree into the single 7,455-line source file that Single had written, left unresolved conflict markers in that file, and ends with snapshots that do not build.

\begin{table}[tbp]\centering\small
\caption{Takeovers from Single's late snapshots: tests passed at the handed-over snapshot (seed) and at the last snapshot within 3\,h, one entry per run. A failed build scores 0. Tests lost to a hung test are rerun and counted (Table~\ref{tab:scoring}).}\label{tab:takeover}
\begin{tabular}{@{}lrll@{}}
\toprule
Task & Seed & New Single & New Multi-Extended \\
\midrule
\task{delta} & 709 / 949 & 719 (mean of two) & 791; 0 (747 at 2.5\,h) \\
\task{lightningcss} & 2,140 / 2,828 & 2,186; 2,238 & 2,216; 2,318 \\
\task{solar} & 1,314 / 1,978 & 1,317; 1,298 & 1,471; 1,423 \\
\bottomrule
\end{tabular}
\end{table}

\paragraph{Why the Multi-Extended takeovers gain more.} The Multi-Extended takeovers gain more than the Single takeovers because their subagents compare more features against the reference program. On \task{lightningcss}, the Multi-Extended takeovers gain 4.5 points on average ($+2.7$ and $+6.3$), and the Single takeovers gain 2.5 points ($+1.6$ and $+3.5$). The list-price cost per point gained is \$74 for the Multi-Extended takeovers against \$16 for the Single takeovers. In 3\,h, the Multi-Extended subagents compare against the reference program 748 and 2,032 times across 15--22 feature areas. In its last 3\,h, the original Single session ran 134 comparisons in 3 areas, and the new Single session ran 33. Their main agent integrates the fixes into functions shared across features and writes about 80\% of the added lines. On \task{solar}, the Multi-Extended takeovers end 140 tests (7.1 points) above the Single takeovers on average, and they cost \$221--346 per run against \$44--67 for the Single takeovers. On \task{delta}, the Single takeovers gain 1.1 points. A new Single session therefore raises Single's score only a little, and more checking against the reference program raises its score by a few points. Across the three tasks, a new Single session gains 1 to 3 points on \task{delta} and \task{lightningcss} and ends below the score of the handed-over code on \task{solar}, while a new Multi-Extended session gains 3 to 9 points where its code builds and spends four to eight times as much.

\paragraph{Fresh against resumed sessions.} In Figure~\ref{fig:context}, the fresh session gains more than the resumed one over the first hour in all six panels. With GPT-5.6 Sol the margin is small, 0.3 to 0.7 points. With GPT-6.1 Sol on \task{lightningcss}, the resumed-session line of Figure~\ref{fig:context} is a repeat of Single-Extended's continuation from the same 8.2\,h snapshot, with a snapshot every 10 minutes. With GPT-6.1 Sol on \task{delta} the fresh session leads by 3.5 points after the first hour, while the resumed session gains almost nothing over the following 12\,h. On the GPT-6.1 Sol \task{delta} panel of Figure~\ref{fig:context}, Single-Extended's code at 8.2\,h scores 81.8. At \$153 of total cost, the two chains of fresh Single sessions end at 91.7 and 89.3, Multi-Extended from scratch stands at 90.1, and Single-Extended, which kept its context, stands at 82.5.

\paragraph{Chains of fresh Single sessions at equal cost.} %
Each new Single takeover of Table~\ref{tab:takeover} was continued as a chain: its final code was handed to a further fresh Single session that had an empty context, a restored copy of the reference program and the same 3\,h continuation prompts. This was repeated for three further sessions, which gives 12\,h of agent time in all. There are two chains per task, six chains in all. Separately, the from-scratch Multi-Extended run on \task{solar} was repeated once more with a clean copy of the reference program, and it ends at 54.4. Its earlier clean repeat ends at 57.4 and the traced run ends at 46.0, and all three are below Single's late score of 63--65. Table~\ref{tab:chain} gives each chain's score after every session and its cumulative list-price cost. %
The equal-cost gaps (last column of Table~\ref{tab:chain}) lie within the 3--8 point difference between two runs of one setup, and the chains spent 42--79\% of a Multi-Extended session's total cost. If we instead read the last Multi-Extended snapshot whose cumulative cost does not exceed the chain's total, the gap has the same sign on every task. Each chain gains 0.5 to 1.5 points per session on average, and no session gains more than 3.9 points. One of \task{solar}'s two chains falls below its seed in the first session and passes it again in the second (snapshots within a session range over 0.2--2.8 points there). With GPT-5.6 Sol on \task{delta}, every single-agent path, whether it is resumed or chained, stays 5.5 to 8.7 points below Multi-Extended from scratch at equal cost. %

\begin{table}[tbp]\centering\small\setlength{\tabcolsep}{4pt}%
\caption{Chains of fresh Single sessions from Single's late snapshot: tests passed at the seed and after each 3\,h session, cumulative list-price cost (USD, session 1 is the new Single takeover of Table~\ref{tab:takeover}), and the new Multi-Extended sessions' mean final score and cost on the same task. Equal cost: Multi-Extended's interpolated score at the chain's total cost.}\label{tab:chain}
\begin{tabular}{@{}l r rrrr r rr r@{}}
\toprule
Chain & Seed & Session 1 & 2 & 3 & 4 & Cost & \shortstack[r]{Multi-Extended\\final (cost)} & \shortstack[r]{Equal\\cost} & \shortstack[r]{Chain $-$\\equal cost} \\
\midrule
\task{delta} 1 & 709 & 719 & 721 & 728 & 734 & \$220 & 791 (\$279) & 765 & $-$2.3 \\
\task{delta} 2 & 709 & 719 & 731 & 739 & 752 & \$196 & & & \\
\task{lightningcss} 1 & 2,140 & 2,186 & 2,202 & 2,253 & 2,282 & \$142 & 2,267 (\$340) & 2,230 & $+$2.3 \\
\task{lightningcss} 2 & 2,140 & 2,238 & 2,267 & 2,294 & 2,307 & \$174 & & & \\
\task{solar} 1 & 1,314 & 1,317 & 1,359 & 1,362 & 1,407 & \$195 & 1,447 (\$283) & 1,434 & $-$0.9 \\
\task{solar} 2 & 1,314 & 1,298 & 1,375 & 1,417 & 1,424 & \$158 & & & \\
\bottomrule
\end{tabular}
\end{table}

\paragraph{Takeovers from Multi-Extended's late snapshots.} The reverse experiment starts from the last snapshot of a Multi-Extended run: \task{delta} at 8.5\,h (868 of 949 tests pass, and the run had gained 1.2 points in its last 3\,h) and \task{solar} at 4\,h (910 of 1,978 tests pass, and its best snapshot, with 997 tests at 1\,h, preceded two main-agent checks in shared stages that broke it). \task{lightningcss} has no such snapshot, because its Multi-Extended run was still gaining when its 4\,h budget ended. Each snapshot is handed to two new Single and two new Multi-Extended sessions for 3\,h, with the reference program's copy restored. At the last snapshot, \task{delta}'s new Single sessions reach 869 and 868, and its new Multi-Extended sessions reach 874 and 877 (the mean gains are $+0.1$ and $+0.8$ points; one Multi-Extended session is rescored with the pager rule and read at its last snapshot, at 2.5\,h, with the same 874 tests, §\ref{app:scoring}). \task{solar}'s new Single sessions reach 1,092 and 1,065, and its new Multi-Extended sessions reach 1,073 and 1,159 (the mean gains are $+8.5$ and $+10.4$). On \task{delta} neither kind of new session gains more than the original run did in its last 3\,h.

On \task{solar} the handed-over state is not slow progress but a regression. In the original run the subagents made no calls between minutes 18 and 139. In that window, the main agent added the two checks that are described below and re-checked both with stored inputs only (31 stored cases at minute 69 and 20 stored fuzz inputs after minute 86, all of which passed). In the same window, 160 tests that passed at minute 60 failed at minute 90, and 157 of them were still failing at the end. These 160 tests turned from pass to fail within the window, whereas the 138 tests of RQ3 are those lost relative to Single. Before the original main agent stopped and the code stopped changing (§\ref{app:scoring}), it had added a newline check to the input stage (between minutes 60 and 70) and a symbol check to semantic validation (between minutes 80 and 90). Both checks reject valid programs, and every re-check that it ran used stored single-line programs with empty function bodies, which reach neither check. All four new sessions wrote their own multi-line, multi-file inputs, triggered the newline check within 4--38 minutes and corrected it, and later corrected the symbol check. Of their gain, 58--77\% consists of tests that the two checks had broken.%

\subsection{Lost Tests and Checking}
\label{app:gaps}

\paragraph{The merge-conflict example.} \task{delta}'s 20 hidden merge-conflict tests cover an empty conflict section, a missing end marker, markers in the wrong order, both sides empty, Unicode content, long lines and a two-way conflict without a common ancestor. Single first compared the feature at about minute 92, and most of its later comparisons loop over the task's bundled example files, which hold one ordinary conflict. At the end, Single passes 5 of the 20 tests and Multi-Extended passes 18.

\paragraph{Features compared by each setup.} Nearly all of Single's lost tests lie in features that Single did compare at some point (Table~\ref{tab:checkcov}), so the losses come from comparing too late or with too few inputs rather than from never comparing. Features and comparisons are defined in §\ref{app:measures}. As in Table~\ref{tab:lost}, a test \emph{lost by Single} fails in Single's final snapshot and passes in Multi-Extended's. In these runs Single's continuation prompt itself asks it to compare against the reference program on new inputs and edge cases (Table~\ref{tab:prompts}).

\begin{table}[tbp]\centering\small
\caption{Comparisons against the reference program, and where Single's lost tests fall. S = Single, M = Multi-Extended.}\label{tab:checkcov}
\begin{tabular}{@{}lrrr@{}}
\toprule
Task & Tests / features & Comparisons, S / M & Lost tests in features S compared \\
\midrule
\task{delta} & 949 / 85 & 1,278 / 6,133 & 144 of 165 \\
\task{lightningcss} & 2,828 / 503 & 340 / 1,295 & 180 of 197 \\
\task{solar} & 1,978 / 98 & 562 / 404 & 30 of 30 \\
\bottomrule
\end{tabular}
\end{table}

On \task{delta}, Single passes 83\% of the tests in features that it first compared within the first hour, 70\% of the tests in features that it first compared between 1 and 4\,h, 67\% of the tests in features that it first compared after 4\,h, and 33\% of the tests in the few features that it never compared. Multi-Extended passes 93\%, 92\%, 88\% and 85\% of the tests in the same groups. On \task{delta} (GPT-5.6 Sol), Multi-Extended had compared half of the 85 features within 12 minutes. Single, with 1,278 comparisons over 20\,h, compared a feature for the first time as late as 11.4\,h. Within the first hour, Multi-Extended had compared 74 of the 85 features and Single had compared 31. Multi-Extended's subagents ran 5,840 of its 6,133 comparisons, and Single reached half of the features only after 1.5\,h and never compared five of them. %
On \task{lightningcss}, the 128 features that only Multi-Extended compared do not lower Single's pass rate (Single passes 76\% of the tests in them against 77\% in the features that Single compared), and the two setups end with nearly the same score. On \task{solar} both setups compare the same features, and Multi-Extended passes fewer of their tests (46\% against 66\%).

\paragraph{Interface cases.} On \task{lua} (GPT-5.6 Sol), Multi-Extended's debug library, which is loaded by name, was never written. With GPT-6.1 Sol on \task{solar}, a subagent called another subagent's module with an argument that the module lacked, and this call made 34 tests fail until a third subagent fixed the module. The fix came 80 minutes after the break. On \task{lua}, GPT-5.6 Sol's Single-Extended ran 4\,h and ends at 94.5, and Multi-Extended ran 2\,h, wrote 13 modules and ends at 90.1.

\paragraph{Main-agent edits to subagents' modules.} In 7 of the 70 runs whose subagents wrote code, the main agent alone made 12 edits to a subagent's module, and each of these edits was followed by the failure of at least five hidden tests that had been passing. Of the 197 tests lost after these edits, 108 never came back.

\paragraph{How Multi-Extended re-checked.} Multi-Extended's main agent re-checks nearly every change to shared code, but mostly with stored inputs. We examine 12 Multi-Extended runs: the three extended runs and nine scored 2\,h Multi-Extended runs on further ProgramBench tasks. A \emph{hand-back} is a subagent's message to the main agent after the subagent wrote product code, and \emph{shared code} is a file through which other features also pass. A hand-back is \emph{checked} if the main agent ran the reference program within 10 minutes after it; a main-agent edit to shared code is \emph{re-checked} if any agent ran the reference program before the next snapshot. A re-check is \emph{broad} if it re-runs a stored set of inputs (a loop over saved cases, a fuzz corpus, a comparison script) and \emph{targeted} if it runs a few hand-written inputs. An \emph{interval of sustained loss} is a pair of consecutive snapshots across which at least 5 tests go from pass to fail and still fail at the next snapshot. Table~\ref{tab:accept} gives the counts.

\begin{table}[tbp]\centering\small
\caption{Checks after subagent hand-backs and after main-agent edits to shared code, Multi-Extended. Loss intervals: intervals of sustained loss in which only the main agent wrote product code, out of all such intervals, with the tests lost in them.}\label{tab:accept}
\resizebox{\linewidth}{!}{\begin{tabular}{@{}lrrrrr@{}}
\toprule
Runs & Hand-backs & Checked within 10\,min & Shared-code edits & Re-checked (broad) & Loss intervals, main agent only \\
\midrule
\task{delta} (8.5\,h) & 425 & 97\% & 179 & 100\% (98\%) & 0 of 1 \\
\task{lightningcss} (4\,h) & 65 & 100\% & 271 & 81\% (48\%) & 8 of 11 \\
\task{solar} (4\,h) & 0 & --- & 142 & 94\% (26\%) & 2 of 2 \\
nine 2\,h runs & 72 & --- & 1,086 & 63--96\% & 14 of 16 \\
all & 562 & 97\% & 1,678 & 85\% (43\%) & 24 of 30 (386 of 454 tests) \\
\bottomrule
\end{tabular}}
\end{table}

Most losses occur after the main agent has re-checked. In 29 of the 30 intervals of sustained loss some agent compared against the reference program afterwards. In 11 of them, a broad re-check reported no difference while hidden tests were failing, and those 11 intervals hold 54\% of the lost tests. The breadth of the re-check after a shared-code edit does not reliably predict whether the edit loses tests. After the main agent's last shared-code edit in an interval, the interval loses tests in 4\% of the cases with a broad re-check (2 of 53), in 11\% of the cases with a targeted re-check (10 of 89) and in 20\% of the cases with no re-check (17 of 83). However, most of this gradient comes from \task{lightningcss}, and with a loss threshold of one test the three rates are 36\%, 40\% and 42\%. The high rate of hand-backs checked within 10 minutes reflects the main agent's routine rather than a response to the hand-back, because the main agent compares against the reference program every 1--5 minutes, and 90\% of hand-backs already had a comparison in the 10 minutes before them. The trajectories do not show whether a given comparison exercised the changed code.

\section{Other Products and Models}
\label{app:rq5}

This section gives the supporting results for §\ref{sec:other} on other products and models: first a replication of RQ1--RQ3 with Opus 5.5 on Claude Code (§\ref{app:ccrepro}), then an analysis of how each model delegates, stops and fares in extended runs (§\ref{app:bymodel}).

\subsection{Replication of RQ1--RQ3 on Claude Code}
\label{app:ccrepro}

We use a small replication to test whether the paper's conclusions hold on another product and model. It runs Opus 5.5 on Claude Code 2.1.283 (effort level high), one run per cell, under the same permission sentence and natural-stop rule as the Codex study. RQ1 uses ten tasks of each benchmark, drawn from the RQ1 random sets by an order that was fixed before any run, and each task is run with Single, Multi, and Multi with Haiku~4.5 as the default subagent model (Table~\ref{tab:ccrepro10}). For RQ2, we run \task{atlas} with every RQ2 setup. For RQ3, we use the extended runs and the six Single 4\,h versus Multi 2\,h pairs on \task{delta} and \task{solar} described in §\ref{app:bymodel}, together with one such pair each on \task{lua} and \task{mdBook} (ProgramBench). Table~\ref{tab:ccrepro} states the result for each conclusion.

\begin{table}[tbp]\centering\footnotesize
\caption{\label{tab:ccrepro10}The RQ1 replication on Claude Code with Opus 5.5, all ten random tasks of each benchmark, one run per cell. Scores are final scores (ProgramBench in \%, NL2Repo as reward $\times$ 100). Cost is the list-price cost in dollars. Subagents is the number of subagents that Multi's main agent started. $^{a}$A lower bound, because the run was cut off at about 2\,h. $^{b}$No final score (see Validity).}
\begin{tabular}{@{}lrrrrrrr@{}}
\toprule
 & \multicolumn{3}{c}{Score} & \multicolumn{3}{c}{Cost} & \\
\cmidrule(lr){2-4}\cmidrule(lr){5-7}
Task & Single & Multi & Haiku & Single & Multi & Haiku & Subagents \\
\midrule
\multicolumn{8}{@{}l}{\emph{ProgramBench}} \\
\task{elfcat} & 100.0 & 99.1 & 100.0 & 8 & 9 & 10 & 0 \\
\task{hex} & 100.0 & 100.0 & 100.0 & 2 & 5 & 2 & 0 \\
\task{quinn} & 98.3 & 99.8 & 98.3 & 20 & 20 & 14 & 1 \\
\task{delta} & 63.4 & 80.7$^{a}$ & 84.3 & 55 & 289 & 141 & 6 \\
\task{solar} & 72.9 & 85.5 & 81.5 & 41 & 107 & 99 & 12 \\
\task{jsonschema} & 99.2 & 99.6 & 99.3 & 32 & 55 & 30 & 4 \\
\task{tuc} & 99.8 & 99.3 & 99.8 & 15 & 34 & 15 & 1 \\
\task{ripsecrets} & 97.4 & 95.7 & 97.5 & 17 & 20 & 50 & 2 \\
\task{lightningcss} & 76.6 & 95.2 & ---$^{b}$ & 53 & 216 & 160 & 13 \\
\task{i3-style} & 99.8 & 99.8 & 99.8 & 12 & 22 & 13 & 2 \\
Mean & 90.7 & 95.5 & 95.6 & & & & \\
\midrule
\multicolumn{8}{@{}l}{\emph{NL2Repo}} \\
\task{verifiers} & 80.7 & 83.0 & 77.8 & 13 & 14 & 8 & 4 \\
\task{frontmatter} & 98.2 & 98.2 & 98.2 & 2 & 3 & 2 & 1 \\
\task{sqlparse} & 98.9 & 99.8 & 99.8 & 13 & 14 & 10 & 3 \\
\task{synthetic} & 78.5 & 77.4 & 73.1 & 8 & 17 & 11 & 5 \\
\task{flask-restful} & 99.7 & 99.7 & 99.7 & 6 & 8 & 6 & 2 \\
\task{pyquery} & 90.5 & 94.6 & 95.9 & 8 & 7 & 9 & 1 \\
\task{markdownify} & 100.0 & 98.8 & 100.0 & 2 & 4 & 3 & 1 \\
\task{markupsafe} & 100.0 & 100.0 & 100.0 & 1 & 2 & 2 & 2 \\
\task{autopep8} & 99.1 & 99.1 & 96.5 & 13 & 22 & 10 & 4 \\
\task{python-pathspec} & 98.3 & 54.6 & 38.7 & 3 & 3 & 3 & 1 \\
Mean & 94.4 & 90.5 & 88.0 & & & & \\
\bottomrule
\end{tabular}
\end{table}

On ProgramBench, Multi ends 4.7 points above Single on average (95.5 against 90.7) at 3.1 times Single's cost. The Haiku mean covers the nine tasks that have a score. On NL2Repo, Single ends highest (94.4, against 90.5 for Multi and 88.0 for Multi with Haiku subagents), and Multi costs 1.4 times as much as Single. The difference on NL2Repo comes from one task, \task{python-pathspec}. Without it, the three setups end at 94.0, 94.5 and 93.4.

\begin{table}[tbp]\centering\footnotesize
\caption{\label{tab:ccrepro}Each conclusion of the paper, checked on Claude Code with Opus 5.5. Each cell is one run. ProgramBench scores are in \%, and NL2Repo scores are the reward $\times$ 100.}
\begin{tabularx}{\linewidth}{@{}>{\raggedright\arraybackslash}p{0.24\linewidth} >{\raggedright\arraybackslash}p{0.16\linewidth} >{\raggedright\arraybackslash}X@{}}
\toprule
Conclusion & Codex, GPT-5.6 Sol & Claude Code, Opus 5.5 \\
\midrule
Main agent delegates under the permission sentence & every run; subagents only probe & mostly on hard tasks: 4--13 subagents on \task{delta}, \task{solar}, \task{jsonschema}, \task{lightningcss} and \task{atlas}; 0--2 on the other six random ProgramBench tasks (Table~\ref{tab:ccrepro10}) \\
\addlinespace
Score rises sooner, ends level (Finding 1) & holds on NL2Repo & holds for 2\,h against 4\,h where it delegates: in the eight 4\,h versus 2\,h pairs (\task{delta} and \task{solar} three times each, \task{lua} and \task{mdBook} once), Multi leads at 2\,h in seven and ties on \task{mdBook}, and Single given twice the time ends level (mean $\Delta C$ +0.8), whereas at the natural stop and with 4\,h against 8\,h Multi ends higher on \task{delta} and \task{solar}; without delegation the setups tie (\task{elfcat} 99.1 vs.\ 100, \task{hex} 100 vs.\ 100) \\
\addlinespace
Lower-cost subagents do not beat Single (RQ1) & holds: Luna subagents end 2.4 (NL2Repo) and 2.2 (ProgramBench) points below Single & holds: over the ten NL2Repo tasks, Haiku subagents end below Single (88.0 vs.\ 94.4, and 93.4 vs.\ 94.0 without \task{python-pathspec}) \\
\addlinespace
Single stops early (Finding 2) & holds & holds: on \task{atlas} Single stops at 61\,min at 81.7; continued to 4\,h it reaches 91.1, told to probe first 91.5 \\
\addlinespace
Default subagents mainly check (Finding 2) & holds & does not apply: its subagents often write product code without being told to \\
\addlinespace
Extra cost of the multi-agent setup (Finding 1) & 3.2 to 3.8$\times$ & about 1.8$\times$ Single's cost over the first 2\,h of the paired runs; 1.7--5.2$\times$ in runs to a natural stop or in extended runs \\
\addlinespace
Growth slows in extended runs (RQ3) & holds & holds: Single holds 74.8 on \task{delta} for almost 5\,h of an 8\,h run \\
\addlinespace
Which setup ends higher (Finding 3) & task-dependent & Multi ends higher on both \task{delta} and \task{solar} \\
\addlinespace
Fresh sessions (Finding 3) & with GPT-5.6 Sol and GPT-6.1 Sol & not run \\
\bottomrule
\end{tabularx}
\end{table}

\paragraph{Readings.} Two conclusions differ from those on Codex (Table~\ref{tab:ccrepro}). First, Opus 5.5's subagents write product code. Second, where the runs go to a natural stop or are extended, its Multi ends higher than Single, at 1.7 to 5.2 times Single's cost. At the natural stop, Multi ends higher by 17.3 points on \task{delta}, which is a lower bound because Multi was cut off at about 2\,h (see Validity; Single reaches the 4\,h cap at 63.4), and by 12.6 points on \task{solar}, where Single stops at 181 minutes. In the extended runs, Multi ends higher by 2.9 and 10.2 points.

\paragraph{Validity.} Each cell is a single run. In Claude Code's non-interactive mode, a session ends when the main agent ends its turn, and this ended subagents that were still running in four cells (two multi-agent cells on \task{atlas}, and the cells on \task{delta} and \task{solar} with Haiku subagents). These four cells were continued from the state at which they were cut, with agent time summed over the continued run. Later cells with subagents keep the session open while subagents run. When we checked all finished cells, we found no other cell with a subagent still running at its end. \task{delta}'s Multi was cut off at about 2\,h, so its score is a lower bound. With Haiku~4.5 as the default subagent model, the main agent may choose Opus 5.5 for an individual subagent, and did so for 17 of 32 subagents on \task{delta} and 5 of 14 on \task{solar}. NL2Repo \task{synthetic}'s Single scores 0 officially because the official test run is killed during one test. When each test file is run in its own process, it scores 0.785, and we use this score. For the other 29 NL2Repo cells, the two ways of scoring agree. \task{jsonschema} is scored as in the Codex study, without stopping at the first failed test (§\ref{app:scoring}). Under the official rule, its Single would score 62.1 instead of 99.2. On \task{lightningcss}, all three runs reached the 4\,h cap. The run with Haiku subagents left no final workspace and no snapshot that builds, so it has no score, and its cost is still counted. On \task{python-pathspec}, the hidden tests import helper functions from a file, \texttt{tests/util.py}, that the benchmark does not supply. Single wrote a version with every helper that the tests import. Multi's main agent wrote a version without two of them, and the run with Haiku subagents wrote no such file, so two and three hidden test files, respectively, fail at import. Cost is priced from the token usage of every main-agent and subagent message, each message counted once, at list prices per million tokens. Opus 5.5 costs \$4 for input and \$20 for output, and Haiku~4.5 costs \$1 and \$5. Cache reads are priced at 0.1 times and cache writes at 1.25 times the input price. Auxiliary calls that Claude Code does not log, which amount to about 2--5\% of its own usage total, are not counted. The takeover of RQ3 was not run on Claude Code.

\paragraph{Early stopping on another harness.}\label{app:dsh}
To check that GPT-5.6 Sol's early stop is a property of the model rather than of Codex, Single and Single-Probe were run once each on the four hard tasks under dsh, an open-source agent harness, with the same model, task text and natural-stop rule. Single stops after 13--20 minutes on every task, as it does under Codex (10--27 minutes). It scores 49.8, 68.9, 61.7 and 63.3 on \task{atlas}, \task{gittype}, \task{broot} and \task{skeema}, against Codex's three-run means of 47.4, 51.6, 59.3 and 68.7. Under dsh, the probe sentence does not help (Single-Probe scores 50.3, 59.5, 53.3 and 66.8), whereas under Codex it raises Single's score by 12--28 points. Because there is only one run per cell, we note this difference but do not interpret it.

\paragraph{Early stopping on NL2Repo, and probe-only subagents.} On NL2Repo's four hard tasks (\task{boltons}, \task{deepdiff}, \task{deslib} and \task{jinja}), GPT-5.6 Sol's Single stops on its own after 26--43 minutes (median 32 minutes). Table~\ref{tab:nl2r2} gives both runs of each setup. NL2Repo has no reference program to probe, so its Single-Probe appends this sentence to the prompt: ``Before you stop, extract every example in the specification and check that your implementation produces exactly the documented result for each one.'' Every Single-Probe run ends below every Single run of its task. One \task{jinja} Single-Probe run scores 0 because it stopped on its own after 51 minutes with two copies of the package in the workspace (\texttt{jinja2/} and \texttt{src/jinja2/}). As a result, the package cannot be installed and no test runs. On Claude Code, a variant of Opus 5.5's Multi whose subagents are told that they may only probe ends between 6 points below and 7 points above the unrestricted Multi, depending on the task (two runs each on \task{atlas}, \task{broot}, \task{gittype} and \task{skeema}).

\begin{table}[tbp]\centering\small
\caption{\label{tab:nl2r2}NL2Repo's four hard tasks with GPT-5.6 Sol. Each cell gives the final score (\%) of each run as run 1 / run 2, with the mean in parentheses. Single-Probe checks every example of the specification before it stops.}
\begin{tabular}{@{}l r rrr@{}}
\toprule
Task & Tests & Single & Multi & Single-Probe \\
\midrule
\task{boltons} & 423 & 69.3 / 70.2 (69.7) & 78.3 / 78.7 (78.5) & 34.0 / 53.0 (43.5) \\
\task{deepdiff} & 970 & 50.9 / 48.9 (49.9) & 58.0 / 61.3 (59.7) & 40.3 / 33.0 (36.6) \\
\task{deslib} & 532 & 48.7 / 48.3 (48.5) & 57.3 / 52.4 (54.9) & 30.6 / 25.4 (28.0) \\
\task{jinja} & 911 & 75.5 / 74.9 (75.2) & 68.1 / 64.7 (66.4) & 0.0 / 56.3 (28.2) \\
\bottomrule
\end{tabular}
\end{table}

\subsection{Delegation, Stopping and Extended Runs by Model}
\label{app:bymodel}

\paragraph{Setup.} All runs in this section use Claude Code 2.1.283 at effort level high, with Opus 5.5, GLM-5.3, Grok 4.6 or MiniMax-M3 as the main agent and the subagent model. The GLM-5.3 runs appear only in this section. The tasks were fixed before any of these runs: \task{delta} and \task{solar}, the two RQ3 tasks on which Codex's Multi-Extended ends higher and lower than Single, respectively; the hard RQ2 task \task{atlas}; and three NL2Repo tasks chosen by a rule written before the runs (the rule picks the tasks whose Codex Single score is closest to the median of the 29 tasks). On \task{delta} and \task{solar}, each model runs Single for 4\,h against Multi for 2\,h, both as extended runs (§\ref{sec:setups}), with 1--3 pairs per model and task. Multi receives the permission sentence of the Codex study, reworded to name Claude Code's Task tool, through which the main agent starts subagents. Delegation is allowed but not required. Table~\ref{tab:rq5} reports $\Delta C$, the final score of Multi at 2\,h minus that of Single at 4\,h.

\begin{table}[htbp]\centering\footnotesize
\caption{Delegation and $\Delta C$ (Multi at 2\,h minus Single at 4\,h, one value per pair of runs) across products and models. All rows except the first are runs on Claude Code. Task calls are the main agent's calls to the Task tool (delegations). Writers are the subagents that wrote product code, shown as writers / all subagents. $^{a}$This value is not $\Delta C$ but the gap between Multi-Extended and Single at an equal 4\,h (§\ref{app:rq3}), over 5 vs.\ 3 runs on \task{delta} and one run each on \task{solar}. $^{b}$There is no delegation in any pair, so the multi-agent run is a repeat of Single. $^{c}$The multi-agent run has no snapshot between 55 minutes and the final workspace, which is scored. $^{d}$In both \task{delta} pairs, the verifier timed out on the final snapshots of at least one setup. $^{e}$Single's build stops producing the executable at 135 minutes, so its final snapshot scores 0. Measured against its last working snapshot (25.0), $\Delta C$ is $-8.1$.}
\label{tab:rq5}
\begin{tabularx}{\linewidth}{@{}>{\raggedright\arraybackslash}p{2.4cm}>{\raggedright\arraybackslash}X>{\raggedleft\arraybackslash}p{2.3cm}>{\raggedleft\arraybackslash}p{1.6cm}@{}}
\toprule
Product / model & Delegation & \task{delta} $\Delta C$ & \task{solar} $\Delta C$ \\
\midrule
Codex, GPT-5.6 Sol\newline (4\,h gap$^{a}$) & Multi: spawns 4--20, subagents write no code in 29 of 30 runs (\task{jsonschema} 0.65); Multi-Extended: subagent share 0--0.93 & +13.8$^{a}$ & $-$15.4$^{a}$ \\
Opus 5.5 & \task{delta} 4--6 calls per pair, 4 / 5 writers in pair 1; \task{solar} 4 calls per pair, 1 / 4 and 0 / 4 writers in pairs 1 and 2 & +14.4, $-$4.2, +1.9\newline (mean +4.0) & +3.8, $-$4.9, $-$4.1\newline (mean $-$1.7) \\
GLM-5.3 & \task{delta} 4 calls, 1 / 4 writers; \task{solar} 0 calls & $-$22.3 & $-$6.3$^{b,c}$ \\
Grok 4.6 & 0 calls on both tasks & n/a$^{d}$ & $-$5.1, $-$9.6$^{b}$ \\
MiniMax-M3 & 0 calls on both tasks & +16.9$^{b,e}$, +2.6$^{b}$ & $-$12.2$^{b}$ \\
\bottomrule
\end{tabularx}
\end{table}

\paragraph{Delegation and direction.} Whether the main agent delegates, and whether its subagents write code, depends on the model. Under the same permission sentence, Opus 5.5 delegates in every run on the hard tasks and delegates rarely on the easier random tasks (Table~\ref{tab:ccrepro}). %
Its $\Delta C$ changes sign across pairs on both tasks. Most of the first \task{delta} pair's gain comes from a low Single run (58.8, against 70.3 and 70.4 in the other two pairs). Codex's \task{solar} gap of $-15.4$ is not reproduced. In all six pairs, Multi leads at 2\,h (by 0.5 to 21.4 points), and Single, when it is given twice the time, reaches about the same final score as Multi on average (on \task{delta}, 66.5 against Multi's 70.5; on \task{solar}, 71.4 against 69.6). This is the RQ1 pattern of an early lead without a higher final score. GLM-5.3 delegates on \task{delta} but not on \task{solar}. On \task{delta}, it loses 22.3 points because its multi-agent build fails to compile for most of the first 75 minutes. Grok 4.6 and MiniMax-M3 never delegate on these two tasks under the permission sentence, so their multi-agent runs are repeats of Single, and their $\Delta C$ measures run-to-run variation. (When the prompt names the Agent tool, Grok delegates three times, only to explore, and gains nothing.)

\paragraph{Stopping.} When and whether a model stops early also depends on the model. Under the deadline instruction of the extended runs, Opus 5.5's Single first stops on its own at 233--236 minutes of its 4\,h in four of its six \task{delta} and \task{solar} runs and does not stop at all in the other two. GLM-5.3's Single never stops, and MiniMax-M3's Single stops after 45--70 minutes. Finding~2 concerns natural-stop runs, which receive no deadline instruction. When it is given only the task prompt, Opus 5.5's Single stops on its own at 61 minutes on \task{atlas} (score 81.7) and at 181 minutes on \task{solar} (score 72.9). However, it reaches the 4\,h cap on \task{delta} (score 63.4) and in a second \task{atlas} run (score 91.7), so its stopping point varies even within a task. On \task{atlas}, continuing it after the early stop raises the score to 91.1, whereas Grok 4.6, which stops only after 137 minutes (score 87.3), gains nothing from continuing (score 87.6). MiniMax-M3 stops on its own after 10--40 minutes on \task{delta}, with scores of 23.9, 23.6 and 27.5 in three runs. A plain ``continue'' adds nothing. The two continued runs score 22.2 and 25.5, and each of them declared itself done ten times in a row. MiniMax-M3 therefore stops early, but it has no room left to improve. GLM-5.3 never stops within the 4\,h cap. Without a deadline instruction, it never writes the required \texttt{compile.sh}, so its natural-stop runs score 0 under the official rule, and we use them only to study stopping behavior.

\paragraph{Extended runs.} When the runs are extended on Claude Code, Opus 5.5's Multi ends above its Single on both \task{delta} and \task{solar}, whereas on Codex Multi-Extended ends below Single on \task{solar} (§\ref{sec:rq3}). We extended the Opus 5.5 pairs to Single for 8\,h and Multi for 4\,h on \task{delta} and \task{solar}, one run per setup. The runs use the deadline instruction and the permission sentence above, are continued after each stop, and receive no parallelism instruction. The multi-agent side is therefore Multi, not Multi-Extended. Every 5-minute snapshot and every final workspace was scored (Table~\ref{tab:opusext}).

\begin{table}[tbp]\centering\footnotesize
\caption{Opus 5.5 extended runs on Claude Code, one run per setup. Each time column gives the score of the last scored snapshot at or before that time. Multi's 4\,h and Single's 8\,h columns show their final snapshots. Last hour: the score of the final snapshot minus the score one hour before it. Subagents: writers / all subagents, and the largest number running at once. Cost: list-price cost of every main-agent and subagent message (prices in §\ref{app:ccrepro}). $^{a}$Single's build first succeeds at 65 minutes.}
\label{tab:opusext}
\begin{tabularx}{\linewidth}{@{}>{\raggedright\arraybackslash}X*{5}{>{\raggedleft\arraybackslash}p{0.8cm}}>{\raggedleft\arraybackslash}p{1.4cm}>{\raggedleft\arraybackslash}p{0.8cm}@{}}
\toprule
Task $\cdot$ setup & 1\,h & 2\,h & 4\,h & 8\,h & Last hour & Subagents & Cost \\
\midrule
\task{delta} $\cdot$ Single & 0.0$^{a}$ & 71.2 & 74.8 & 78.4 & +3.6 & --- & \$94 \\
\task{delta} $\cdot$ Multi & 64.5 & 80.4 & \textbf{81.3} & --- & +0.1 & 28 / 28, 14 & \$217 \\
\task{solar} $\cdot$ Single & 63.3 & 69.3 & 77.9 & 81.3 & +0.2 & --- & \$102 \\
\task{solar} $\cdot$ Multi & 67.6 & 76.1 & \textbf{91.5} & --- & +1.6 & 9 / 12, 4 & \$167 \\
\bottomrule
\end{tabularx}
\end{table}

The scores of all four runs rise slowly late in the run. On \task{delta}, Single reaches 74.8 at about 3\,h and holds it for the next five hours while writing about 2,800 more lines (one snapshot at 6.6\,h fails to build and scores 0). Its last 3.6 points come from a three-line fix to the hunk-header label, which it made in the continuation turns after its first stop. Multi on \task{delta} gains less than one point in its second two hours. On \task{solar}, Single gains less than one point in its last 2\,h. Multi's score jumps by 10 points at 130 minutes, when the main agent merges two stages that its subagents wrote, the type checker and the AST dump (about 4,700 lines), and Multi gains 1.6 points in its last hour. %
Both Multi runs delegate. On \task{delta}, every one of the 28 subagents is given a file or module to write. On \task{solar}, three subagents only document the reference program's behavior and nine write code. Under the 8\,h deadline instruction, Opus 5.5's Single first stops on its own at 463 minutes on \task{delta} and 478 minutes on \task{solar}, against 233--236 minutes under the 4\,h instruction, so its stopping point depends on the stated budget. Multi first stops at 225 and 234 minutes of its 4\,h.

\paragraph{File count.} Among the main-agent models, only GPT-5.6 Sol, and GPT-6 Sol when it works alone, leave the product in one or two files (Table~\ref{tab:structure}). When they work alone on Claude Code, the other models port the program into modules in the original language. Opus 5.5 leaves 13 to 25 Rust or Go files (median 15). GLM-5.3 leaves 7 to 13 files, Grok 4.6 leaves 8 to 15, and MiniMax-M3 leaves a median of 7. Under Codex, Grok 4.6 leaves 8 to 11 files. How much of the code is shared therefore depends on the model as well as on the task.

\paragraph{NL2Repo.} On the three NL2Repo tasks repeated with other models (§\ref{app:rq5}), Opus 5.5 scores 0.93--1.00 in every setup, and MiniMax-M3's multi-agent runs score below its Single runs on the two tasks where both are scored (with one run of each setup).

\section{Case Studies}\label{app:cases}

The body of the paper reports what each setup scores. This appendix shows what the agents do in single runs. We chose the cases to illustrate the mechanisms of RQ3, and we do not use them to estimate how often these mechanisms occur.

\subsection{How a Single Agent Gets Stuck}\label{app:case-stuck}

GPT-5.6 Sol's Single-Extended on \task{delta} passes 706 of 949 tests at 10.5\,h and 714 at 20\,h (Figure~\ref{fig:plateau}). Its slow progress comes from a target that it chose rather than from code that it cannot change. At 6\,h, its output on all 34 examples that are shipped with the task matches the reference program's output byte for byte, and from then on the agent has no external signal. It then turns to inputs that it generates itself. At 12 to 13\,h, it writes a random diff generator with 500 cases over eight mock languages. In its last 6\,h, 79\% of its edit--test cycles fix such generated cases, against 0.3\% in its first 6\,h. In these last 6\,h, a cycle gains 0.006 hidden tests on average, against 0.69 per cycle in the first 6\,h. Over these last 9.5\,h, only 12 hidden tests change state, and 218 of the 235 tests that are still failing at the end produce the same output throughout.

The remaining failures cluster by feature, and the agent no longer visits those features. Among its last 590 cycles, at least 0, 0 and 5 run inputs on submodules, navigation and blame, respectively, the features that hold many of the failures. A separate 4.3\,h multi-agent run on the same task passes 143 of these 235 tests. Over the same stretch of score, its main agent hands out 36 new tasks by feature area, so its subagents keep spreading over unfinished features. Context loss is not the cause. The session compacts its context 15 times, and each summary contains an exact pointer to the generated case that the agent is working on.

\subsection{Why a Fresh Session Starts Faster}\label{app:case-fresh}

\begin{figure}[tbp]\centering
  \includegraphics[width=0.95\linewidth]{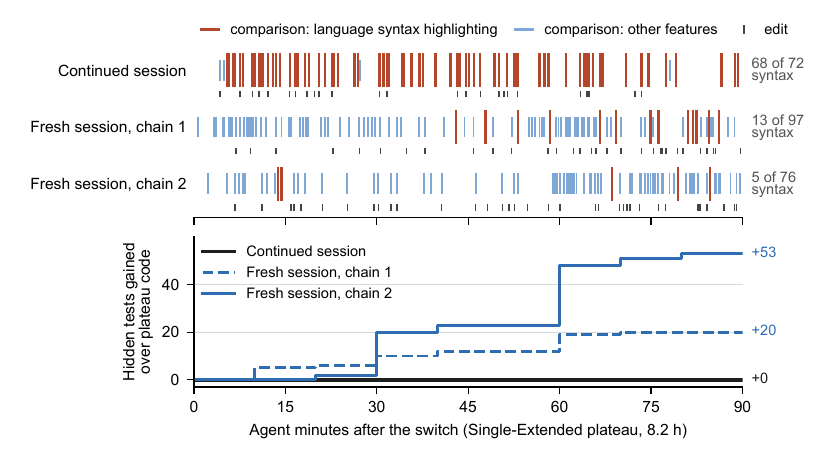}
  \caption{\label{fig:casefresh}\textbf{After 8.2\,h, the resumed session keeps checking syntax highlighting and gains no test; fresh sessions from the same code compare many features and gain tests within the first half hour.} GPT-6.1 Sol, \task{delta}, the first 1.5\,h after the restart at 8.2\,h. Top: one lane per session; each tall tick is a comparison against the reference program (dark: language syntax highlighting; light: other features), and each small mark is an edit. Bottom: hidden tests gained over the code at 8.2\,h (best so far, 10-minute snapshots).}\end{figure}

With GPT-6.1 Sol on \task{delta}, the resumed session had narrowed its work to a target of its own making, and the fresh sessions went back to covering many features (Figure~\ref{fig:casefresh}). At 8.2\,h, when the chain starts, most of Single-Extended's own checks matched the reference program, and its context named one open item: stored syntax-highlighting outputs of the reference program for many languages and themes. In the following hour, 52 of the resumed session's 55 comparisons replay these outputs. Its code changes fall in 7 functions, almost all of them in the highlighter, and the set of hidden tests that it passes does not change (776 of 949). Only one of the 173 failing tests concerns language highlighting. The two fresh sessions start from the same code, read the implementation notes and the manual, and write new comparison scripts that cover options, diffs and styles. In their first hour, they compare 37 and 22 features, against 6 for the resumed session. They change 23 and 26 functions and pass 19 and 48 more tests, in clusters such as \texttt{-{}-show-themes} (19 tests). The narrowing began before the session compacted its context 2.5\,min after the restart. In the 3\,h before, 34 of its 83 comparisons were already on syntax highlighting. This is one task with one resumed run. On \task{lightningcss} and \task{solar}, the continued session keeps gaining, and the early lead of a fresh session is about half a point.

\subsection{Code Structure}\label{app:case-structure}

\begin{figure}[tbp]\centering
  \includegraphics[width=1.0\linewidth]{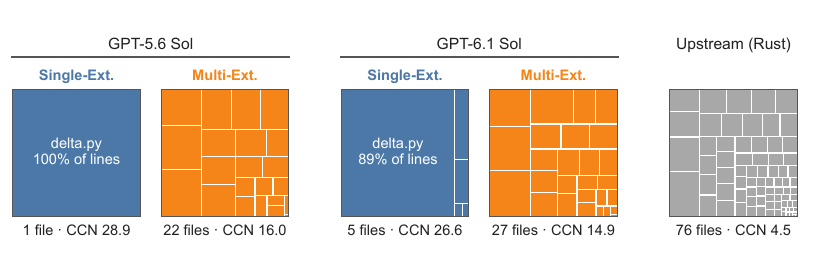}
  \caption{\label{fig:casestructdelta}\textbf{On \task{delta}, Single-Extended keeps its program in one file or nearly one, while Multi-Extended splits it into 22 to 27 files with simpler functions, closer to the upstream source's 76 files.} Each square is one program's product code at the snapshots of Table~\ref{tab:structure_long}, and each rectangle is one source file, whose area is the file's share of the program's lines. All squares have equal size regardless of the number of lines. Below each square: number of files and mean cyclomatic complexity (CCN) per function. Generated data tables are included (23\% of GPT-6.1 Sol Multi-Extended's lines).}\end{figure}

On \task{delta}, we observe a single agent that keeps its program in one file and a multi-agent setup that divides it into modules (Figure~\ref{fig:casestructdelta}). With GPT-5.6 Sol, Single-Extended wrote all 7,728 lines in one file, while Multi-Extended wrote 22 files, the largest of which holds 12\% of its lines. With GPT-6.1 Sol, Single-Extended wrote five files but kept 89\% of its lines in one of them. Multi-Extended wrote 27 files, and again none of them holds more than 12\% of its lines. The upstream source spreads 13,320 lines of Rust over 76 files, and none of them holds more than 9\%. Function complexity is ordered in the same way. The mean cyclomatic complexity is 28.9 for Single-Extended against 16.0 for Multi-Extended with GPT-5.6 Sol, and 26.6 against 14.9 with GPT-6.1 Sol. It is 4.5 in the upstream source. Subagents wrote 94\% and 95\% of Multi-Extended's lines. With GPT-6.1 Sol, a quarter of its lines are data tables that subagents generated from the reference program's output. With GPT-5.6 Sol the two runs differ in cost, but Multi-Extended already had its 22 files when its cost reached Single-Extended's total cost of \$260. Both setups remain far from the upstream source.

\begin{figure}[tbp]\centering
  \includegraphics[width=1.0\linewidth]{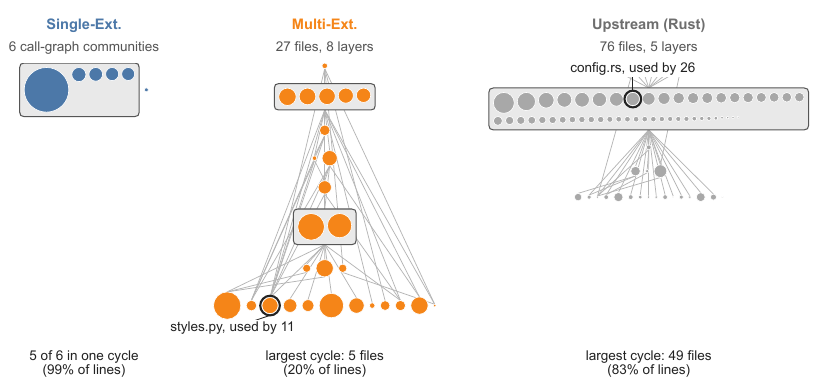}
  \caption{\label{fig:casedeps}\textbf{On \task{delta}, the subagents' files form a layered, almost acyclic dependency graph, while most of the upstream source's files form one cycle.} The figure shows the GPT-6.1 Sol snapshots of Table~\ref{tab:structure_long} and the upstream Rust program. Nodes are call-graph communities for Single-Extended, 89\% of whose code is in one file, and files for Multi-Extended and the upstream source. Node area is proportional to the number of lines. A node that uses another node is drawn above it. Mutually dependent nodes (a cycle) are boxed and drawn as one block, and edges are drawn between blocks only. The ringed node is the file that the most other files use.}\end{figure}

Finer modularity does not mean a dependency structure closer to the upstream source's (Figure~\ref{fig:casedeps}). With GPT-6.1 Sol, Multi-Extended's 27 files form eight layers with two small cycles, one of five files and one of two files. With GPT-5.6 Sol, its 21 Python files form six layers with no cycle. In both runs, about a fifth of the references between top-level functions and classes cross files, and the most used file is a style module that 11 or 12 other files import. The upstream source is the reverse. Of its 76 files, 49 form one cycle around \texttt{config.rs}, and these 49 files hold 83\% of its lines. \texttt{config.rs} is used by 26 files, and about two thirds of the upstream source's references cross files. Single-Extended keeps 89\% to 100\% of its lines in one file. Inside that file, five of the six call-graph communities with GPT-6.1 Sol, and all seven with GPT-5.6 Sol, depend on each other. Rust and Python programs differ in file size, so these counts are indicative only.

\begin{figure}[tbp]\centering
  \includegraphics[width=1.0\linewidth]{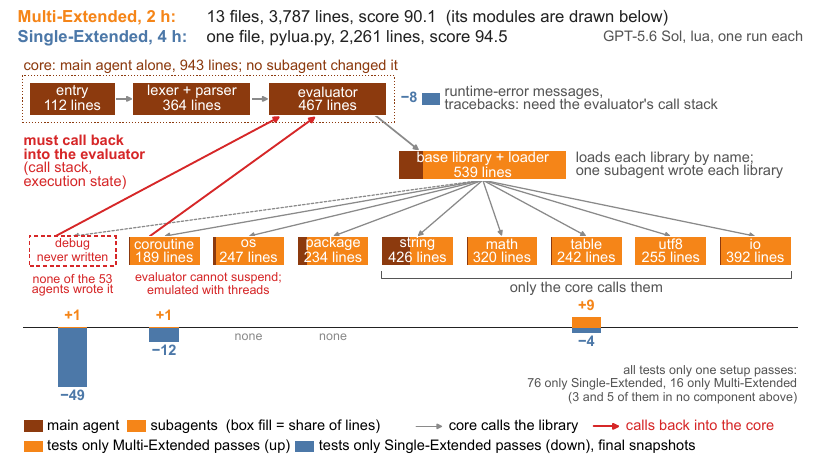}
  \caption{\label{fig:caselua}\textbf{On lua, Multi-Extended's modules gain tests where only the core calls them and lose tests where they must call back into a core that no agent owned.} GPT-5.6 Sol, Multi-Extended (2\,h) against Single-Extended (4\,h), one run each. The diagram shows Multi-Extended's final code. The main agent alone wrote the core (entry, lexer and parser, evaluator), and the loader registers nine standard libraries, each written by one subagent. Each box is a module, filled in proportion to the shares of its lines that were last written by the main agent (dark) and by subagents (orange). The debug library, which is loaded by name, was never written (dashed). Grey arrows: the core calls the library. Red arrows: the library must call back into the evaluator for its call stack, which the main agent did not provide and no subagent changed. Bars: tests that only one setup passes in its final snapshot, up for Multi-Extended, down for Single-Extended. The five libraries that only the core calls are counted together.}\end{figure}

Figure~\ref{fig:caselua} shows both sides of modularity on \task{lua}, the case discussed in §\ref{sec:rq3-lose}. Multi-Extended's main agent wrote the core alone and gave each standard library to one subagent. Multi-Extended gains tests on the libraries that only the core calls. It loses tests on the debug and coroutine libraries, which must call back into the evaluator for its call stack, because no agent owned that interface.

\section{Benchmark Task Prompts}
\label{app:prompts}
Table~\ref{tab:prompts} gives every text a setup adds; this section gives, verbatim, the benchmark task prompt that every run receives before those texts, as recorded in the run records.

Every ProgramBench run receives the benchmark's own task prompt, identical in every run:
\begin{Verbatim}[fontsize=\scriptsize,breaklines=true,breakanywhere=true]
You are a helpful assistant that can interact with a computer.

This is a reverse-engineering benchmark. You are given a compiled binary and its documentation.
Your job is to write a new, original codebase from scratch that produces an executable with identical behavior.
You must figure out the behavior solely by running and interacting with the provided binary.
Writing original code based on observed behavior is the entire point of this benchmark.

Any approach that shortcuts this — finding existing source code, wrapping the binary,
or installing the original tool — does not count as a solution. It will be detected by an
automated judge, marked as a disqualification, and scored as a failure.
The trace will be posted to an online wall of shame.

<RULES>
## What is not allowed

### 1. Obtaining source code

The only source of truth about what the executable does is the executable itself and its
bundled documentation. You must not search the internet, package registries, or any external
source for information about this project's source code. Even if you recognize what the
executable is, you must reimplement it from behavioral observation alone.

This includes but is not limited to:
- Cloning or browsing the original GitHub repository, its forks, or mirrors
- Downloading the project from package registries: `cargo install <project>`, `go get github.com/<org>/<project>`, `pip install <project>`, `apt-get source <project>`, `npm install <project>`, etc.
- Fetching source tarballs from project websites (e.g., `curl https://lua.org/ftp/lua-5.5.0.tar.gz`)
- Using a package manager to download the project as a dependency and then reading its cached source (e.g., navigating into `~/.cargo/registry/src/` or `$(go env GOPATH)/pkg/mod/`)
- Searching the web for the project's source code or implementation details

### 2. Wrapping or reusing the original binary

Your submission must be a genuine reimplementation. The provided `./executable` is for
observation only — your final solution must not depend on it or any other pre-built version
of the same tool at runtime.

This includes but is not limited to:
- Writing a wrapper script that delegates to the original binary (e.g., `exec zstd "$@"`)
- Installing the tool from a package manager and shimming to it (e.g., `apt-get install nnn && cp $(which nnn) ./executable`)
- Writing a `compile.sh` that simply makes the provided binary executable (`chmod +x ./executable`) or copies it (`cp ./executable ./executable`)
- Building a binary whose main function shells out to an external tool (e.g., `Command::new("miniserve").args(args).exec()`)
- Re-linking prebuilt `.o` object files found in the workspace without writing new source code

### 3. Binary analysis of the provided executable

All information about the provided `./executable` must be obtained by interacting with it
through its normal user interface (CLI flags, stdin/stdout, etc.).
- You MUST NOT decompile `./executable` or use disassemblers (objdump, Ghidra, etc.) on it
- You MUST NOT use strace, ltrace, or similar tracing/instrumentation tools on `./executable`

Note: this restriction applies ONLY to the provided `./executable`. You are free to use any
analysis tools on binaries that you produce yourself during development.

## What IS allowed

- Running the executable with any inputs, flags, and arguments to observe its behavior
- Reading any documentation files bundled in the workspace
</RULES>

## Task context

We want to write the source code for a given executable.
The executable is located at `./executable` in the workspace root.

You also have access to the existing documentation.

## Your task

Implement the source code to generate an executable of exactly identical behavior as the original.

No project-specific dependencies are pre-installed.
You do NOT have access to the internet.
**IMPORTANT**: Make sure that the executable(s) and everything else that is an artifact is not committed, i.e., is in your `.gitignore` file.
Finally, commit your changes.

Make sure that you have a `./compile.sh` file that produces an executable `./executable` in the workspace root.
`compile.sh` should be executable and should install any dependencies needed to compile the executable.
If your compile.sh fails to compile on a fresh checkout, your task has failed.

## Important: This is a reverse-engineering benchmark

Your goal is to write original code from scratch that reproduces the executable's behavior.
The only way to learn what the executable does is to run it and read its bundled documentation.

Any attempt to obtain source code — whether successful or not — or to wrap/reuse the
provided binary will be detected by an automated judge, disqualified, and scored as zero.
See the full rules in the system prompt above. Key points:

- Do NOT search the internet, clone repos, or download the project from any package registry
- Do NOT wrap, shim, or delegate to the provided `./executable` or any installed version of the same tool
- Do NOT decompile the provided `./executable` or use strace/ltrace on it (analyzing your own binaries is fine)
- You SHOULD extensively test the executable to understand its behavior before writing code.
  If you are dealing with a TUI, tmux/libtmux has been installed to help you test/inspect/it.

## Recommended Workflow

1. Explore all documentation files
2. Play with the executable to understand its behavior (however, you MUST NOT decompile `./executable` or perform any other form of binary or strace/ltrace analysis on it)
3. Write the source code to implement the behavior

Submit your changes and finish your work by issuing the following command: `echo COMPLETE_TASK_AND_SUBMIT_FINAL_OUTPUT`.
Do not combine it with any other command. <important>After this command, you cannot continue working on this task.</important>
\end{Verbatim}

Every NL2Repo run receives the task instruction of the benchmark harness, identical for all 30 tasks except the project name (shown here for \task{aiofiles}); the specification itself is the file \texttt{/workspace/start.md}:
\begin{Verbatim}[fontsize=\scriptsize,breaklines=true,breakanywhere=true]
Build the complete Python repository described in `/workspace/start.md`.

Start from the specification and create all required source and packaging files in the empty
workspace. Use only tests and checks that you author yourself. The agent environment contains no
reference source or hidden verifier tests. Do not seek or copy the original `aiofiles` project.

The independent verifier runs after the agent exits. It replaces submission-authored tests and
package metadata with canonical evaluator-only copies before scoring the submitted source.
\end{Verbatim}

\end{document}